\documentclass[epj]{svjour}
\usepackage{amsmath}                    
\usepackage{amssymb}                    
\usepackage{graphicx}
\usepackage{color}

\begin{document}

\title{Dijet angular distributions at $\sqrt{s}=14$ TeV}

\author{Nele Boelaert
\and
Torsten \AA kesson}
\institute{Department of Physics, Div. of Experimental High-Energy Physics, S\"olvegatan 14, SE-223 62 Lund, Sweden}
\abstract{
We present a Monte Carlo study of dijet angular distributions at $\sqrt{s}=14$ TeV. First we perform a next-to-leading order QCD study; we calculate the distributions in four different bins of dijet invariant mass using different Monte Carlo programs and different jet algorithms, and we also investigate the systematic uncertainties coming from the choice of the parton distribution functions and the renormalization and factorization scales. 
In the second part of this paper, we present the effects on the distributions coming from a model including gravitational scattering and black hole formation in a world with large extra dimensions. Assuming a 25$\%$ systematic uncertainty, we report a discovery potential for the mass bin $1 < M_{jj} < 2$ TeV at $10$ pb$^{-1}$ integrated luminosity.
}
\maketitle{}

\section{Introduction}
\label{intro}

Jet production is the most dominant hard process in hadron collision experiments. While jets are background for many new physics searches, jets can also be used as a signal, both for probing QCD and for physics beyond the Standard Model. Because of their rich abundance, many jet studies can be performed with little integrated luminosity. 

The dijet angular distribution between the two hardest jets in the event has shown to be a very useful measurement \cite{UA1,UA2,PhysRevLett.77.5336,PhysRevLett.80.666}; at low integrated luminosity it is a good tool to probe QCD, while with more statistics, a search for new physics, such as effects coming from large extra dimensions, becomes possible.

This paper is about dijet angular distributions at $\sqrt{s}=14$ TeV.  First we will perform a QCD study; we will calculate the distributions up to NLO and make an estimate of the systematic uncertainties. In the second part we will study the distributions in a scenario with large extra dimensions, including gravitational scattering and black hole formation.


\section{Dijet angular distributions}
\label{Dist}
In proton-proton collisions, QCD will mainly manifest itself in $t$-channel gluon exchanges, giving rise to a cross section which is approximately inversely proportional to $\hat{t}^2$:
\begin{equation}
 \frac{d\hat{\sigma}}{d\hat{t}} \propto\frac{\alpha_s^2}{\hat{t}^2},
\label{cross}
\end{equation}
with $\hat{t}=-\frac{\hat{s}}{2}(1-\cos\hat{\theta})$, $\hat{\theta}$ being the scattering angle in the center of mass frame of the colliding partons. This corresponds to Rutherford scattering for constant $\hat{s}$, i.e.~a scattering process that peaks at small angles: $d\hat{\sigma}/d(\cos\hat{\theta}) \propto \sin^{-4}(\hat{\theta}/2)$.

Experimentally, the study of the angular behavior is done using the variable $\chi$, which is defined as $\chi=\text{exp}(|\eta_1-\eta_2|)=\text{exp}(2|\eta^*|)$, with $\eta_1$ and $\eta_2$ the pseudorapidities of the two hardest jets in the event and $\eta^*=\frac{1}{2}(\eta_1-\eta_2)$. At lowest order, i.e.~for a $2 \rightarrow 2$ process, the pseudorapidity of the two jets (partons) in the center of mass frame is given by $\pm \eta^*$ and can be related to the scattering angle $\hat{\theta} = \arccos(\tanh\eta^*)$, and therefore also the following expression for $\chi$ as a function of $\hat{\theta}$ holds:
 \begin{equation}
 \chi  =\frac{1+|\cos\hat{\theta}|}{1-|\cos\hat{\theta}|}   \sim \frac{1}{1-|\cos\hat{\theta}|} \propto \frac{\hat{s}}{\hat{t}}
\end{equation}
Using the approximation that $\chi\propto\hat{s}/\hat{t} $ and keeping $\hat{s}$ fixed, the cross section in equation (\ref{cross}) can now be rewritten as a function of $\chi$ and turns out to be approximately constant:
\begin{equation}
\frac{\text{d}\hat{\sigma}}{\text{d}\chi} \propto \frac{\alpha_s^2}{\hat{s}}  \quad\text{(}\hat{s}\text{ fixed)}
\end{equation}
To get the result at the hadron level, the above cross section needs to be multiplied with the parton distribution functions and integrated over the momentum fractions:
\begin{equation}
\frac{\text{d}\sigma}{\text{d}\chi} =  \int \text{d} x_1\int \text{d} x_2f_1(x_1,Q^2)f_2(x_2,Q^2)\frac{\text{d}\hat{\sigma}}{\text{d}\chi} 
\end{equation}
Keeping  $\hat{s}$ fixed implies that to a good approximation, also the product of the two parton distribution functions is fixed (up to logarithmic scaling variations with $Q^2$), so that after integration, the cross section $d\sigma/d\chi$ becomes constant as well. 
This means that, binned in the dijet invariant mass $M_{jj}$ ($\sqrt{\hat{s}}=M_{jj}$ at LO), the angular distribution $\text{d}\sigma/\text{d}\chi$ versus $\chi$ is approximately flat for QCD.  
On the other hand, new physics is very often an $s$-channel resonance or a process characterized by a certain mass threshold. These processes are usually more isotropic than QCD, with a cross section that tends to be flat in $\cos\hat{\theta}$, which causes the dijet angular distributions to peak at low $\chi$. 
Therefore, in order to gain most knowledge from dijet angular distributions, it is necessary to study them in bins of $M_{jj}$.

The following four mass bins were chosen for $\sqrt{s}$=14 TeV: $0.5 < M_{jj}< 1$ TeV, $1 < M_{jj} < 2$ TeV, $2 < M_{jj} < 3$ TeV and 3 TeV $< M_{jj}$.
\par
In experimental conditions the measurable phase space will limit the maximum $\chi$ value one can reach. If $\eta_{\mathrm{max}}$ is the maximum pseudorapidity for which jets are still fully seen by the detector, then the following two orthogonal selection cuts need to be made in order to measure the angular distribution without acceptance losses coming from the limited pseudorapidity range. 
\begin{gather}
|\eta_1-\eta_2|<2\eta_{\mathrm{max}}-c \label{v1}\\
|\eta_1+\eta_2|<c \label{v2}
\end{gather}
For example the ATLAS \cite{atlas} calorimeters can measure jets fully up to $\eta_{\mathrm{max}} \sim$ 4. The choice of $c$ is made by considering the trade-off between having enough statistics passing selection cut (\ref{v2}) and the measurable $\chi$ range, and for our study $c=1.5$ has turned out to be a reasonable compromise. With this choice for the value of $c$ and with $\eta_{\mathrm{max}}$ = 4, the angular distributions can be measured up to $\chi_{\mathrm{max}}\sim$ 600.
In case we are only interested in measuring up to $\chi_{\mathrm{max}}\sim$ 100 (e.g. for new physics searches), we can limit ourselves to $\eta_{\mathrm{max}}$ = 3.1.
\par
At lowest order, the relation between $M_{jj}$, $\chi$ and the transverse momentum $p_T$ is the following:
\begin{equation}
M_{jj}= p_T(\sqrt{\chi}+1/\sqrt{\chi})  \label{v3}
\end{equation}
The selection cuts on dijet mass ($M_{jj,\mathrm{min}}<M_{jj}<M_{jj,\mathrm{max}}$) and $\chi$ ($\chi<\chi_{\mathrm{max}}$), will determine the minimum $p_T$ the two highest jets in the event need to have in order to pass the selection cuts: 
\begin{equation}
p_{T,\mathrm{min,LO}} = \frac{M_{jj,\mathrm{min}}}{(\sqrt{\chi_{\mathrm{max}}}+1/\sqrt{\chi_{\mathrm{max}}})}  \label{v4}
\end{equation}
NLO contributions will lower the minimum transverse momentum with a factor $\sqrt2$, as a consequence of the fact that the second highest jet in the event can never have a transverse momentum less than half the transverse momentum of the highest jet. So, up to NLO:
\begin{equation}
p_{T,\mathrm{min}} = \frac{M_{jj,\mathrm{min}}}{\sqrt{2}(\sqrt{\chi_{\mathrm{max}}}+1/\sqrt{\chi_{\mathrm{max}}})}  \label{v5}
\end{equation}
Equation (\ref{v5}) will be used as a $p_T$-selection cut. Table \ref{tab:1} summarizes this for the different mass bins and for two different values of $\chi_{\mathrm{max}}$.

\begin{table}
\caption{Values of $p_{T,\mathrm{min}}$ for 4 different mass bins and 2 values of $\chi_{\mathrm{max}}$. }
\begin{center}
\begin{tabular}{c c c}
\hline  \hline
Mass bin  	& $\chi_{\mathrm{max}}=100$ 	& $\chi_{\mathrm{max} }=600$ \\
		& $\text{pT}_{\mathrm{min}}$ (GeV)  	& $\text{pT}_{\mathrm{min}}$ (GeV)  \\
\hline 
 $0.5 < M_{jj}< 1$ TeV	& 35			&	14	\\
 $1 < M_{jj} < 2$ TeV	& 70			&  	28 	\\
 $2 < M_{jj} < 3$ TeV	& 140			&	57	\\
 3 TeV $< M_{jj}$	& 210			&	86	\\\hline
\end{tabular}
\end{center}
\label{tab:1}
\end{table}

\section{QCD calculations}
\label{QCD}

Two programs are available for NLO jet calculations: JETRAD \cite{giele-1993-403} and NLOJET++ \cite{Nagy:2003tz}. The programs use a conceptually very different approach; JETRAD uses the phase space slicing technique, while NLOJET++ applies the Catani-Seymour dipole subtraction scheme \cite{catani-1998-510} with some modifications introduced because of computational reasons.

While the considerations in the previous section are at the parton level, experiments have to deal with jets made of hadrons, which are the result of three major steps: hard interaction, parton showering and hadronization. Jets carry the memory of the hard interaction and a good jet-finding algorithm can exhibit this information in an infrared and collinear safe way.  
For many years, the longitudinally-invariant $k_T$-clustering algorithm has shown to be infrared and collinear safe \cite{Catani1993187,PhysRevD.48.3160}, and it will be used in this study. Recently, SISCone \cite{siscone}, a practical infrared safe cone algorithm, has become available and we will also show results using this algorithm. 
Although seeded cone algorithms are not infrared stable, we will also present results with a seeded cone algorithm because they are still frequently used in experiments. More precisely, we will use a seeded, iterative cone with progressive removal that comes with JETRAD and was brought into use by the CERN UA1 collaboration \cite{UA1-Cone}.
NLOJET++ comes with an exact seedless cone algorithm which was first proposed in \cite{RunII} and is finding all stable cones in a given configuration, ensuring infrared safety. We will also show results using this algorithm. 

Note that the final state for a NLO order calculation at the matrix element level contains at most three partons and that we consider only those events with at least two jets. Therefore most problems emerging from infrared and collinear instabilities are absent in our calculations. 
The major difference between the previously mentioned cone algorithms is that the JETRAD seeded cone clusters two nearby partons if their separation in $(\eta ,\phi)$ is less than the cone radius $R$, while the NLOJET++ seedless cone and SISCone only do so if the cone containing both partons is stable, i.e.~if their separation is smaller than $R(1+z)$, with $z= p_{T,2}/ p_{T,1}$ and  $p_{T,2}< p_{T,1}$ \cite{siscone}.

\par
\begin{figure*}
\includegraphics[angle=-90,width=0.9\textwidth]{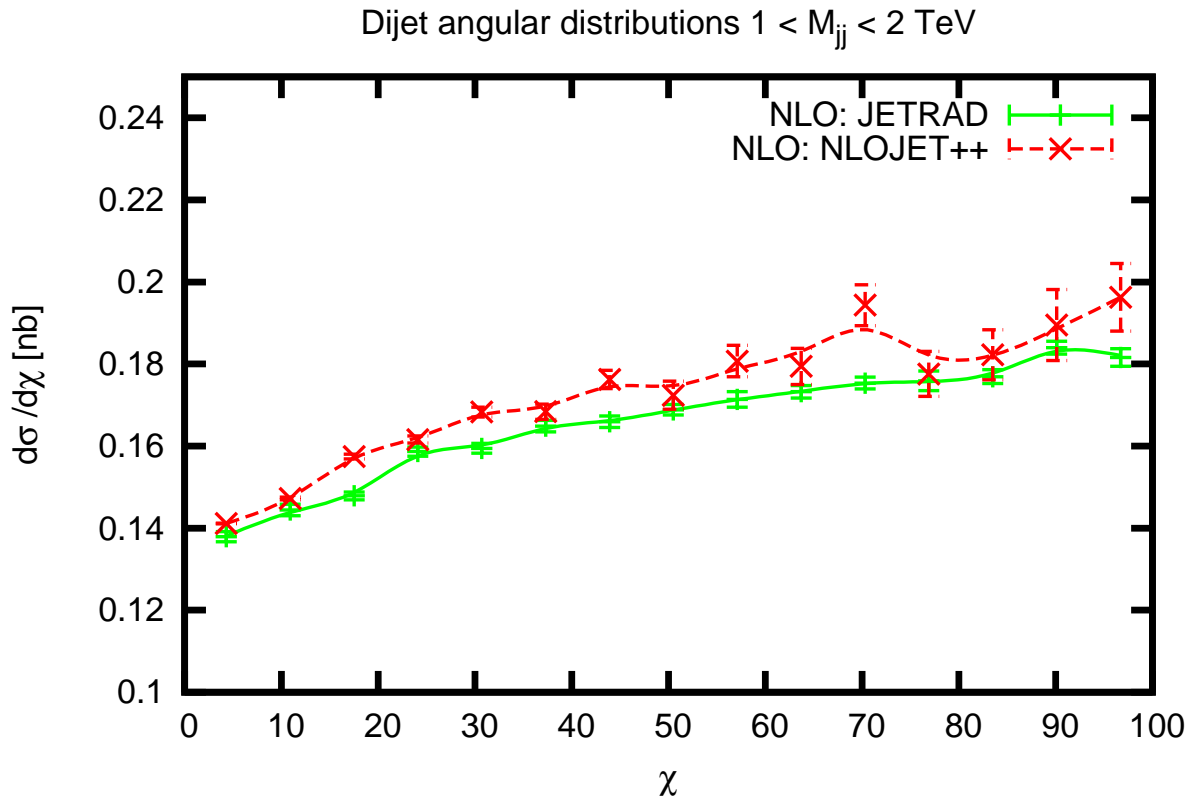}
\caption{Comparing JETRAD with NLOJET++ for the mass bin $1 < M_{jj} < 2$ TeV. The calculations are done at NLO using an inclusive $k_T$ algorithm with R = 1.0.}
\label{nlojet_jetrad}
\end{figure*}
\par
\begin{figure*}
\includegraphics[angle=-90,width=0.9\textwidth]{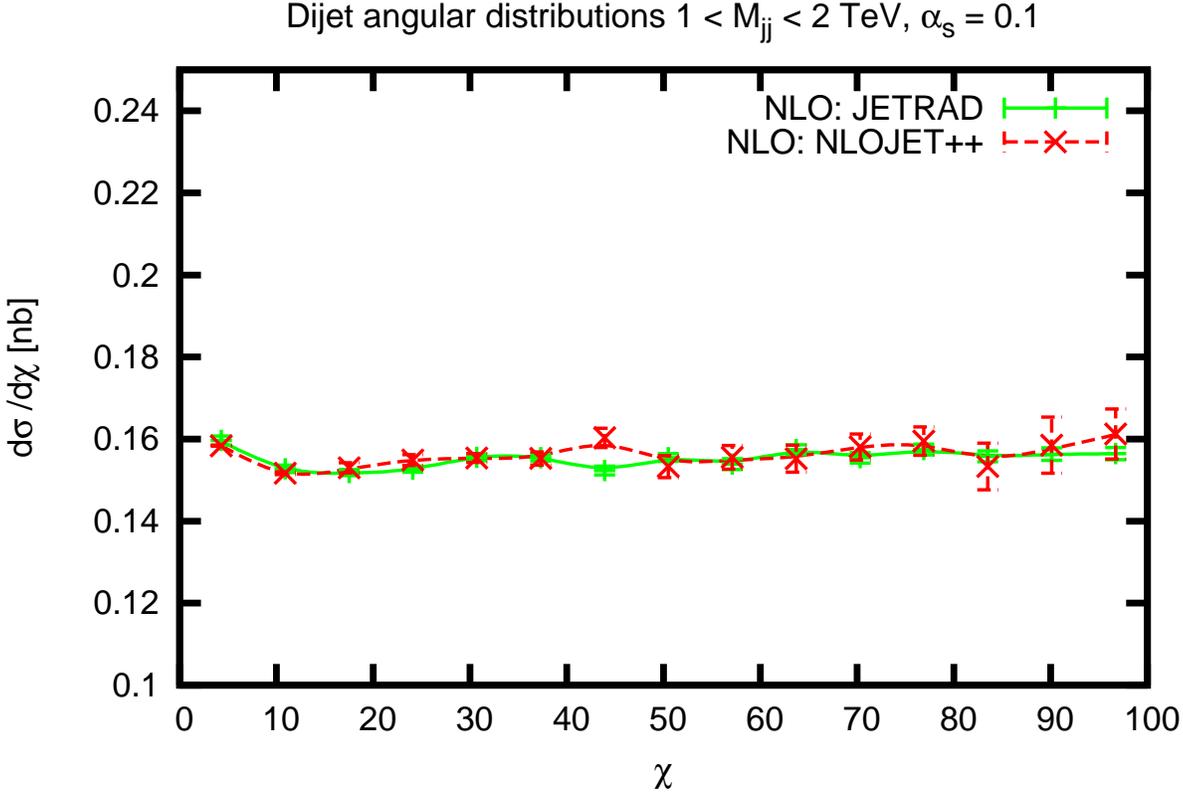}
\caption{Comparing JETRAD with NLOJET++ for the mass bin $1 < M_{jj} < 2$ TeV and for $\alpha_s=0.1$. The calculations are done at NLO using an inclusive $k_T$ algorithm with R = 1.0.}
\label{alphas}
\end{figure*}

Figure \ref{nlojet_jetrad} shows the angular distribution for the mass bin $1 < M_{jj} < 2$ TeV, calculated both with JETRAD and NLOJET++, using an inclusive $k_T$ algorithm with radius parameter R = 1.0. 
JETRAD uses a different parametrization of the strong coupling constant than NLOJET++, which explains the difference between the curves to a large extent. To illustrate this, we show in figure \ref{alphas} the angular distributions with $\alpha_s$ kept constant at 0.1. 
\par
NLOJET++ (coded in C++) has the advantage over JETRAD (coded in fortran) that it can be combined with more modern jet algorithms, such as SISCone and an exact seedless cone. But it has the disadvantage that the angular distributions have a statistical error that is not homogeneous over the whole $\chi$ range, but is instead increasing with $\chi$, which is not the case for a JETRAD calculation, as can be seen from figure \ref{nlojet_jetrad}. 
\par
\begin{figure*}
\begin{center}
\includegraphics[width=1.0\textwidth]{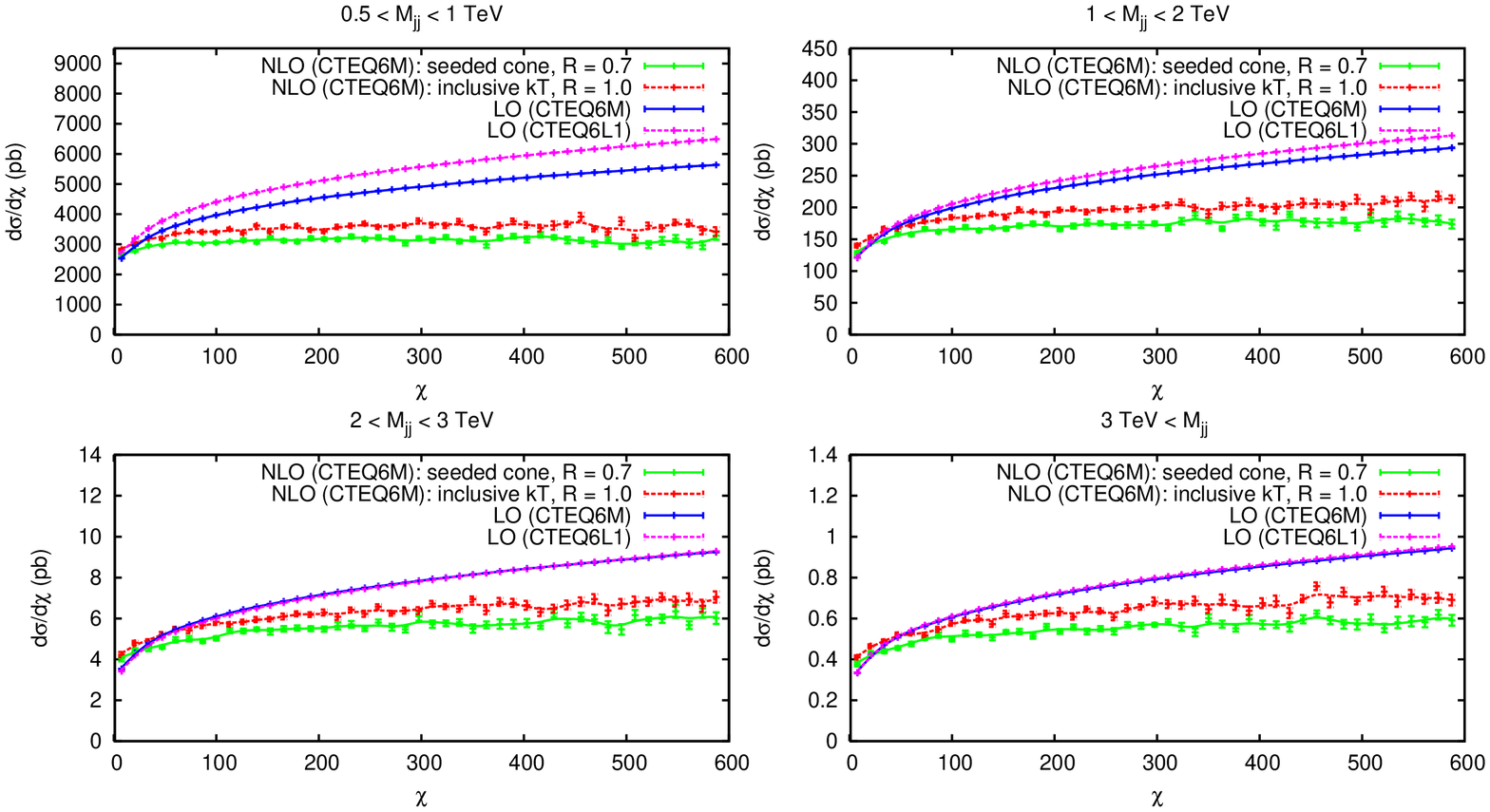}
\caption{LO and NLO angular distributions calculated with JETRAD for 4 different mass bins. The NLO calculations are done with two different jet algorithms: the JETRAD seeded cone algorithm with R=0.7, and an inclusive $k_T$ algorithm  with R = 1.0. All NLO curves are calculated with the CTEQ6M PDF, while the LO distributions are calculated with both the CTEQ6M and the CTEQ6L1 PDF. In the bottom two plots, the LO CTEQ6L1 and LO CTEQ6M curves are overlapping.}
\label{jetrad}
\end{center}
\end{figure*}
Figure \ref{jetrad} compares calculations at the Born (lowest order) level with next-to-leading order calculations, done with JETRAD, for the four different mass bins and for $\chi < 600$. ALL NLO distributions have been calculated with the next-to-leading order CTEQ6M parton distribution function (PDF) \cite{pumplin-2002-0207}, while the leading order distributions have been calculated both with the CTEQ6M PDF and with the leading order CTEQ6L1 PDF.
Two different jet algorithms are used for the NLO calculations; the JETRAD seeded cone algorithm with radius R=0.7, and an inclusive $k_T$ algorithm  with R = 1.0. Note that in a LO parton level calculation, the outgoing partons are back-to-back, so that a jet algorithm is redundant. The NLO angular distributions with the two different jet algorithms tend to have the same shape, but differ in absolute normalization.  

\begin{figure*}
\begin{center}
\includegraphics[angle=-90,width=0.8\textwidth]{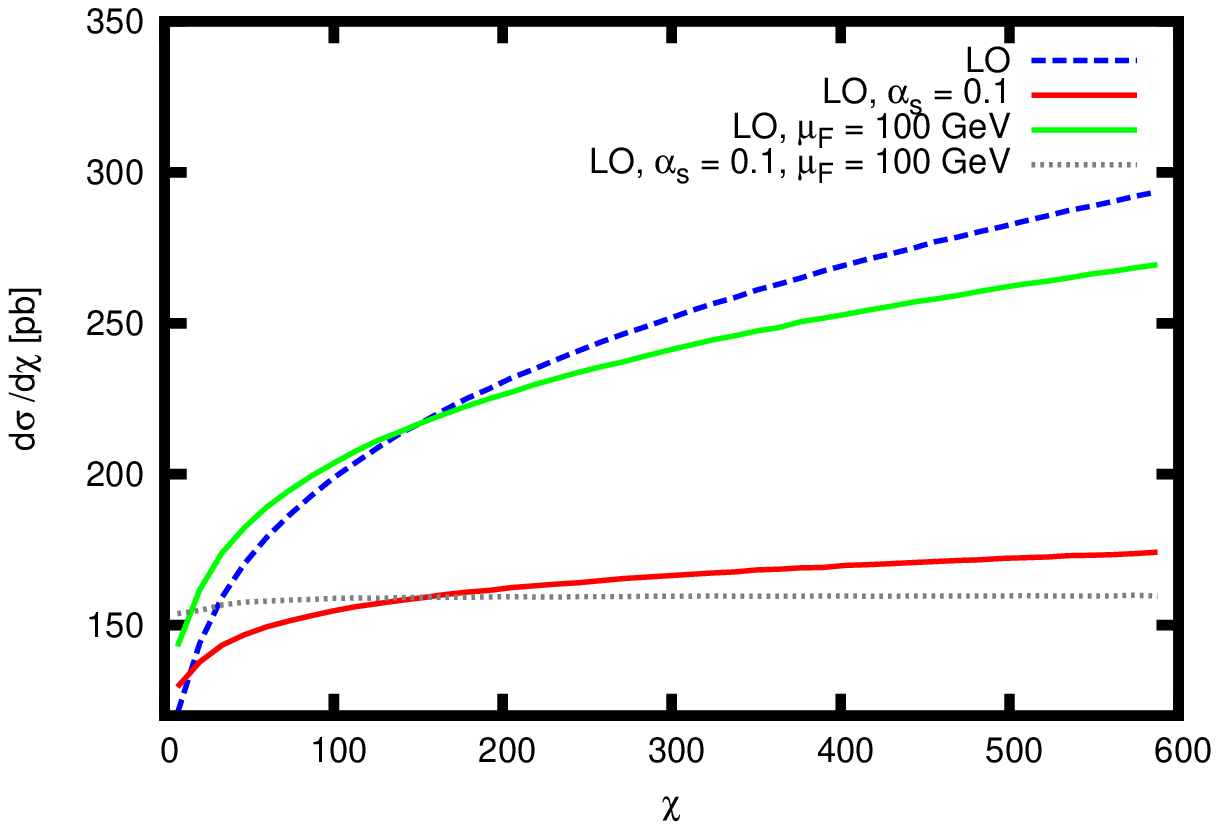}
\caption{Illustrating the influence of the running of $\alpha_s$ and the variation of $\mu_F$ on LO distributions.}
\label{Born}
\end{center}
\end{figure*}

From this figure, we also note that the NLO calculations are flatter than the Born calculations, especially at high $\chi$ values, which can be explained to a large extent by the fact that the running of $\alpha_s$ with $p_T$ (or equivalently $\chi$) has more effect on a LO than a NLO calculation. 
To illustrate this, we have plotted in figure \ref{Born} four different LO calculations in the mass bin $1 < M_{jj} < 2$ TeV; the dashed blue curve is the same Born calculation as the one presented in the top right plot of figure \ref{jetrad}. For this calculation, the value of $\alpha_s$ and of the factorization scale ($\mu_F$) have been varied according to the $p_T$ of the hardest jet. The other three curves show what happens if $\alpha_s$ and/or $\mu_F$ have been kept constant. The grey dotted curve is a calculation with $\alpha_s = 0.1$ and $\mu_F=100$ GeV and is much flatter compared to the blue one. The red curve has been calculated with $\alpha_s = 0.1$ and no running of $\mu_F$, and the green one with $\mu_F=100$ GeV and no running of $\alpha_s$. There is a much bigger effect on the distributions from keeping $\alpha_s$ fixed than from  $\mu_F$. 
\begin{figure*}
\begin{center}
\includegraphics[angle=-90,width=0.8\textwidth]{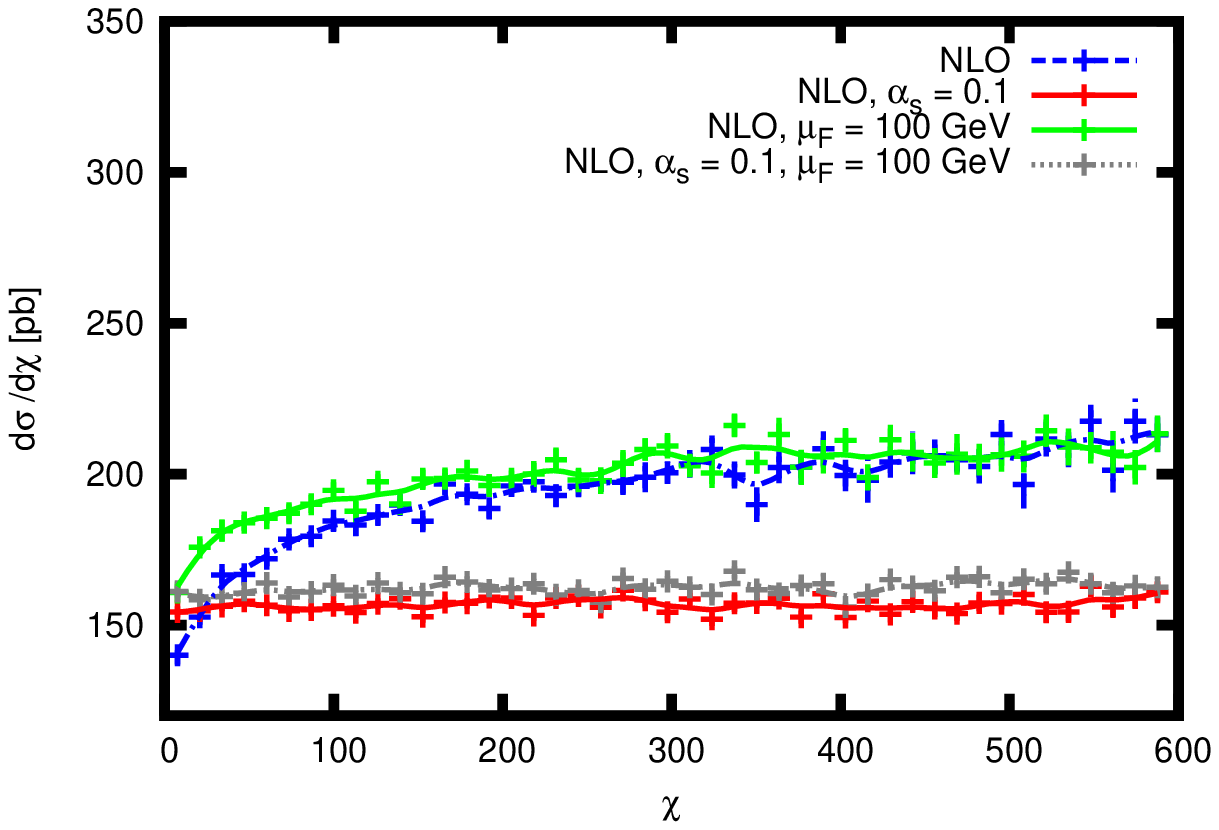}
\caption{Illustrating the influence of the running of $\alpha_s$ and the variation of $\mu_F$ on NLO distributions. The calculations are done with JETRAD and an inclusive $k_T$ algorithm with R = 1.0.}
\label{NLO}
\end{center}
\end{figure*}
\par
But these are observations at the Born level. At NLO, the sensitivity due to scale variations is reduced as can be observed in figure \ref{NLO}. This is because perturbation theory tells us that an all-order calculation should not depend on the renormalization scale at all, and therefore, compared to the LO calculation, a NLO calculation is more stable against the running of $\alpha_s$. Furthermore, a NLO order calculation is using a NLO expansion of $\alpha_s$, and the running of $\alpha_s$ at NLO is less pronounced than at LO. 
Figures \ref{Born} and \ref{NLO} clearly illustrate the need for a NLO order calculation with the running of $\alpha_s$ and $\mu_F$ enabled. 
\par
\begin{figure*}
\begin{center}
\includegraphics[width=1.0\textwidth]{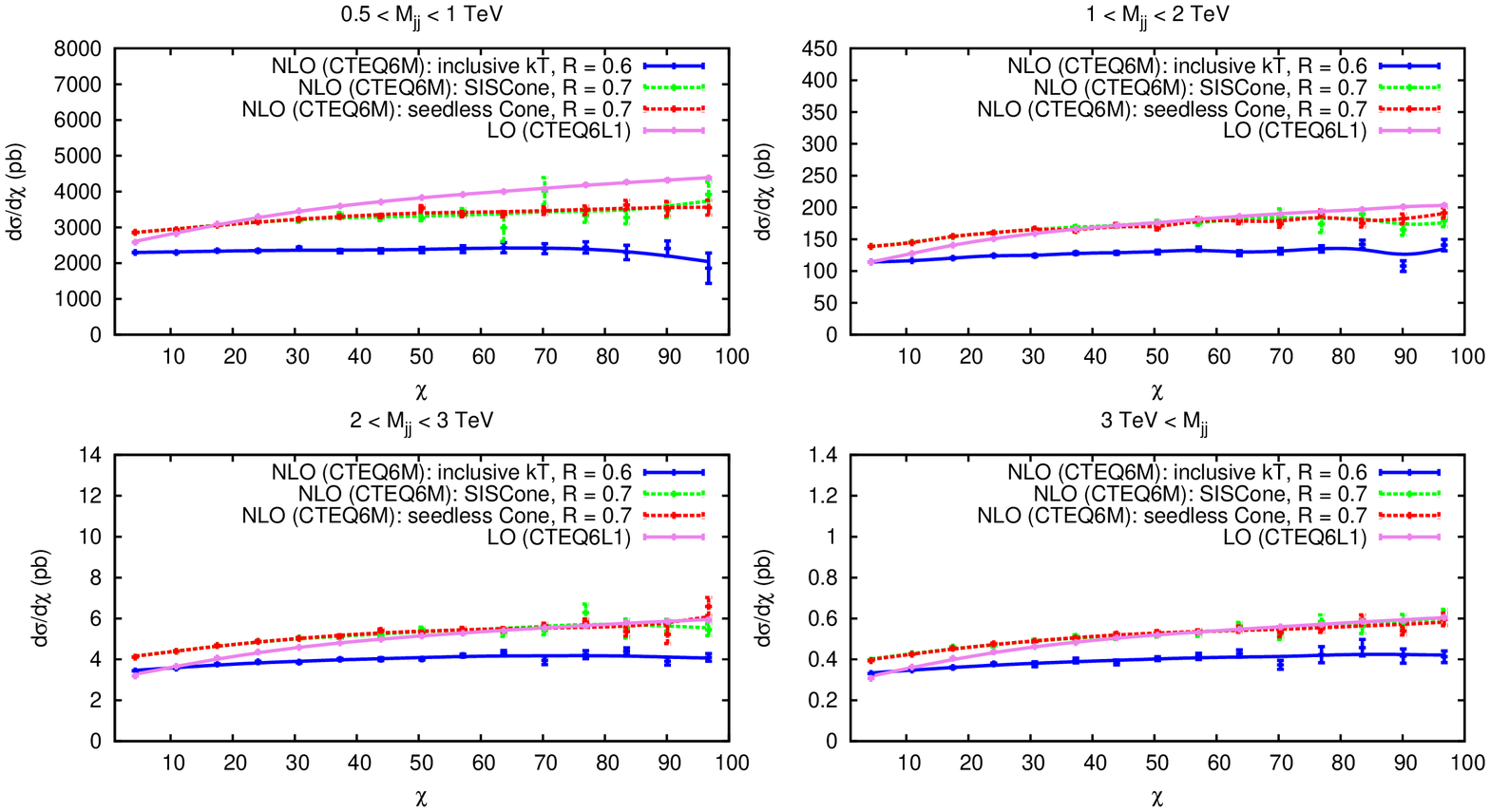}
\caption{LO and NLO angular distributions calculated with NLOJET++ for 4 different mass bins. The NLO calculations are done with three different jet algorithms: an inclusive $k_T$ algorithm  with R = 0.6, SISCone with R = 0.7, and the NLOJET++ seedless cone algorithm with R = 0.7. }
\label{nlojet}
\end{center}
\end{figure*}
\begin{figure*}
\begin{center}
\includegraphics[width=1.0\textwidth]{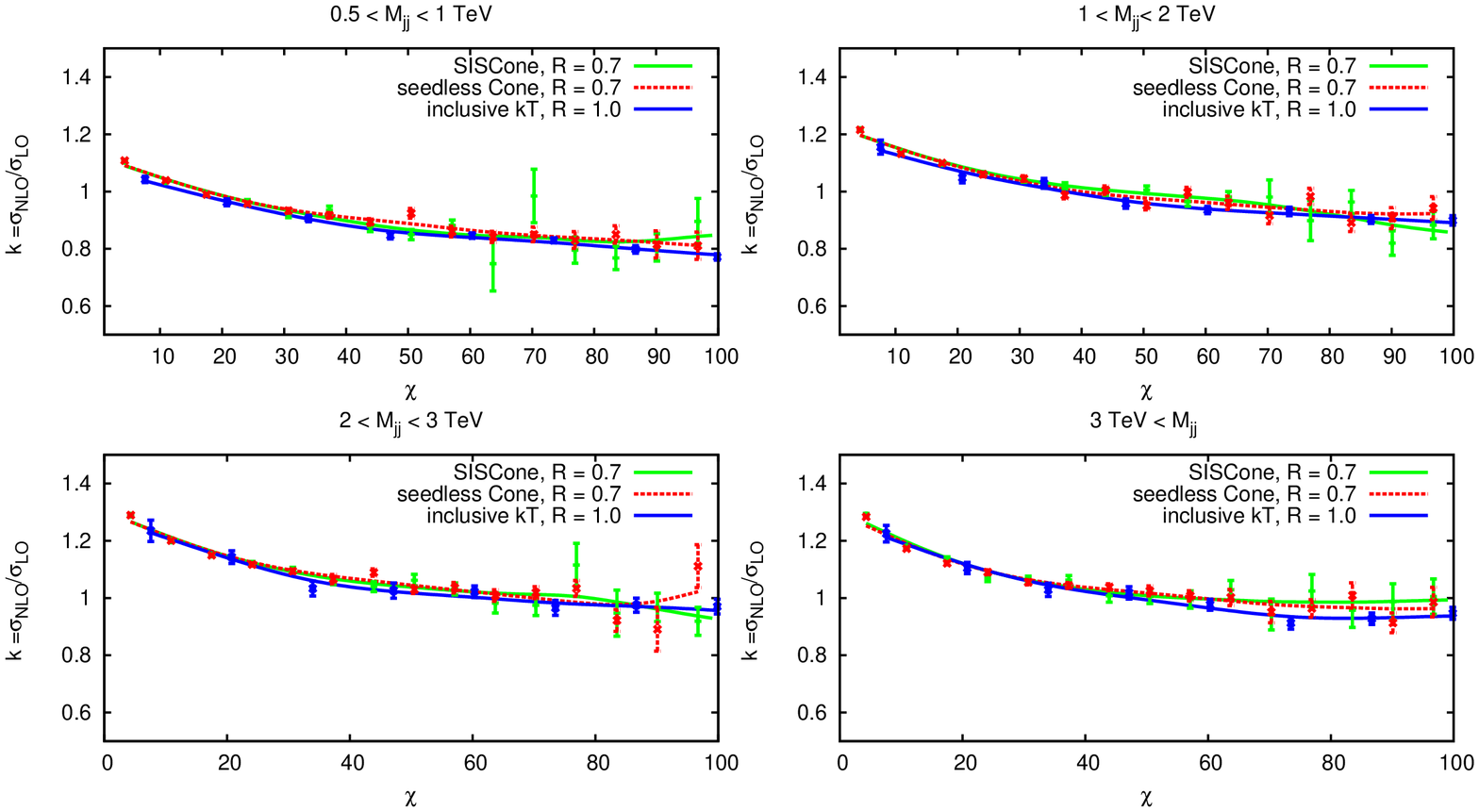}
\caption{Ratio of $\sigma_{\mathrm{NLO}}$(CTEQ6M) over $\sigma_{\mathrm{LO}}$(CTEQ6L1) for 4 different mass bins calculated with NLOJET++.}
\label{Mjj}
\end{center}
\end{figure*}
\par
This is also confirmed by calculations done with NLOJET++. In figure \ref{nlojet} we present the angular distributions, done with NLOJET++, for the same mass bins, using the NLOJET++ seedless cone algorithm with R = 0.7 and overlap 0.5, the SISCone algorithm with R=0.7 and overlap 0.75, and an inclusive $k_T$ algorithm with R = 0.6. We show the distributions for $\chi < 100$ only and, compared to the big variations between the LO and NLO calculations observed at high $\chi$ ($\chi>100$) in figure \ref{jetrad}, the difference between LO and NLO is much less at small values of $\chi$.  
\par
In figure \ref{Mjj} we show the ratio of the NLO cross section calculated with CTEQ6M over the LO one calculated with CTEQ6L1, for $\chi < 100$ and for different jet algorithms, a quantity which is often called the k-factor in literature. The k-factor varies around 1 with a deviation of 30$\%$ at the most. 
\par
The calculations above have been done with the CTEQ6M and CTEQ6L1 parton distribution functions and with a normalization and factorization scale chosen to be the transverse momentum of the hardest jet. Uncertainties coming from parton distribution functions (PDFs) and the choice of renormalization ($\mu_R$) and factorization ($\mu_F$) scale will contribute to a systematic error. We will investigate them using JETRAD with an inclusive $k_T$ algorithm with R = 1.0. The exact scale and PDF uncertainties may vary with different jet algorithms and cone sizes. 
\begin{figure*}
\begin{center}
\includegraphics[angle=-90,width=0.8\textwidth]{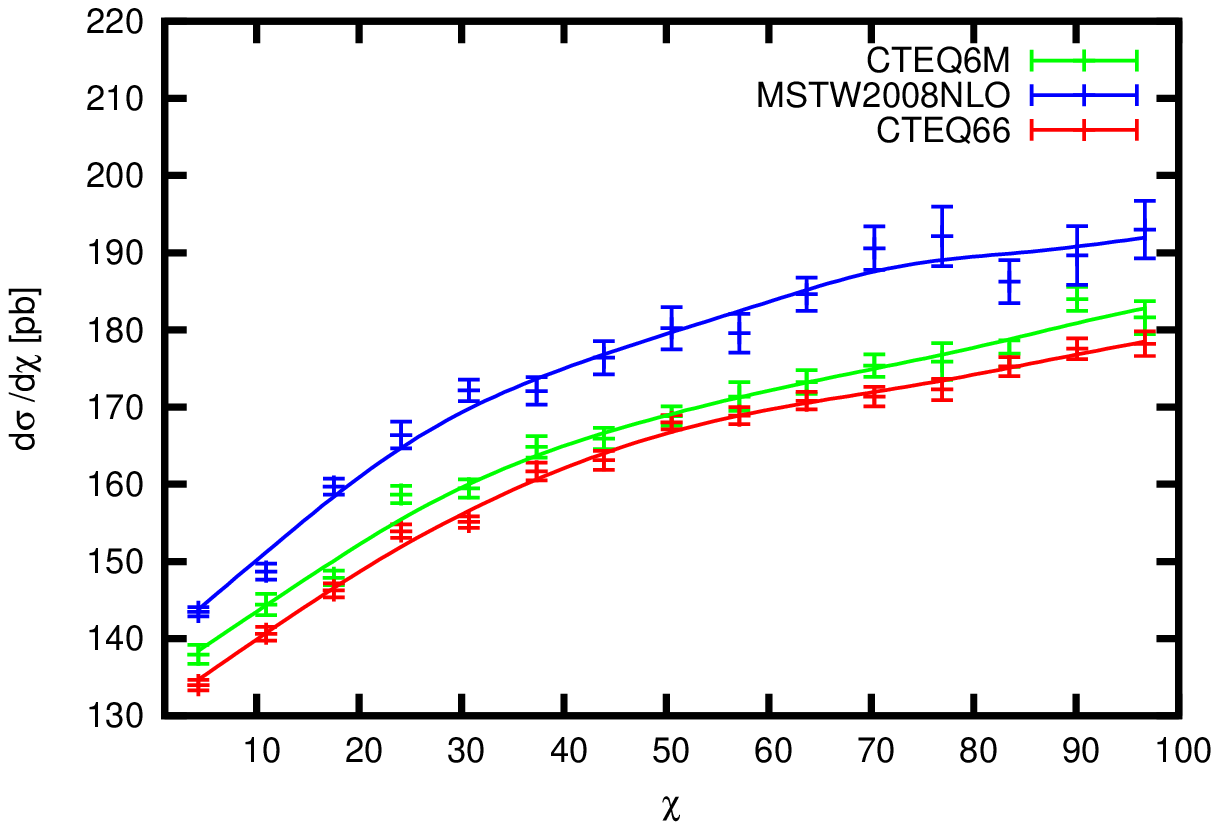}
\caption{Angular distributions with different PDF-sets for the mass bin $1 < M_{jj} < 2$ TeV. The calculations are done with JETRAD and an inclusive $k_T$ algorithm with R = 1.0.}
\label{pdfs}
\end{center}
\end{figure*}

\begin{figure*}
\begin{center}
\includegraphics[angle=-90,width=0.7\textwidth]{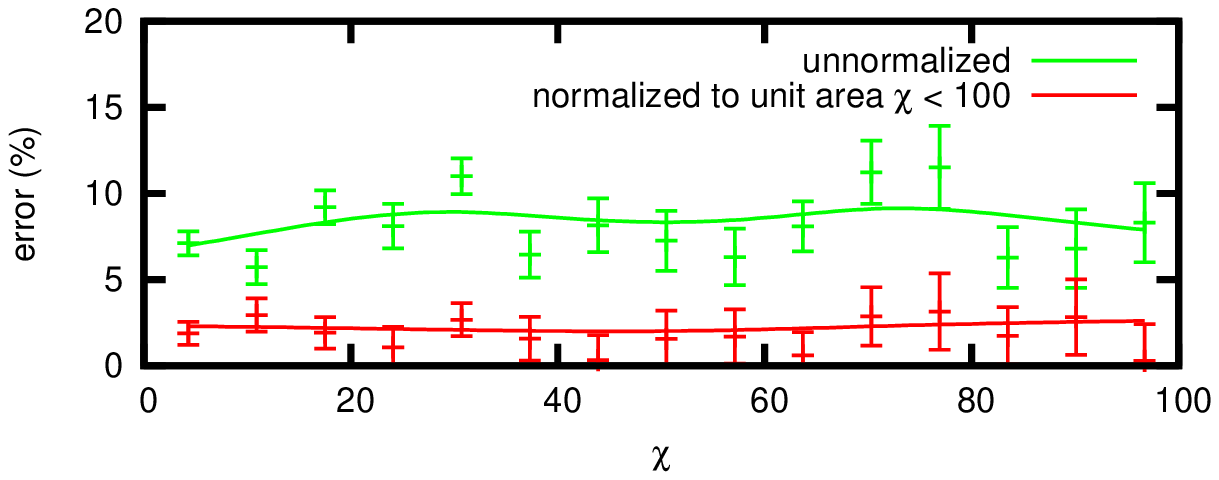}
\caption{Difference caused by the use of different PDF-sets on both normalized and unnormalized dijet angular distributions for the mass bin $1 < M_{jj} < 2$ TeV.}
\label{jetrad_pdfs}
\end{center}
\end{figure*}

Figure \ref{pdfs} shows the angular distributions for the mass bin $1 < M_{jj} < 2$ TeV for three different PDF-sets, namely CTEQ6M, CTEQ66 \cite{nadolsky-2008} and MSTW2008NLO \cite{martin-2009}. The distributions differ mainly in absolute normalization, and less in shape. Normalizing the distributions to unit area make them differ no more than $3\%$ over the whole $\chi$ range, as can be seen in figure \ref{jetrad_pdfs}.

\begin{figure*}
\begin{center}
\includegraphics[angle=-90,width=0.8\textwidth]{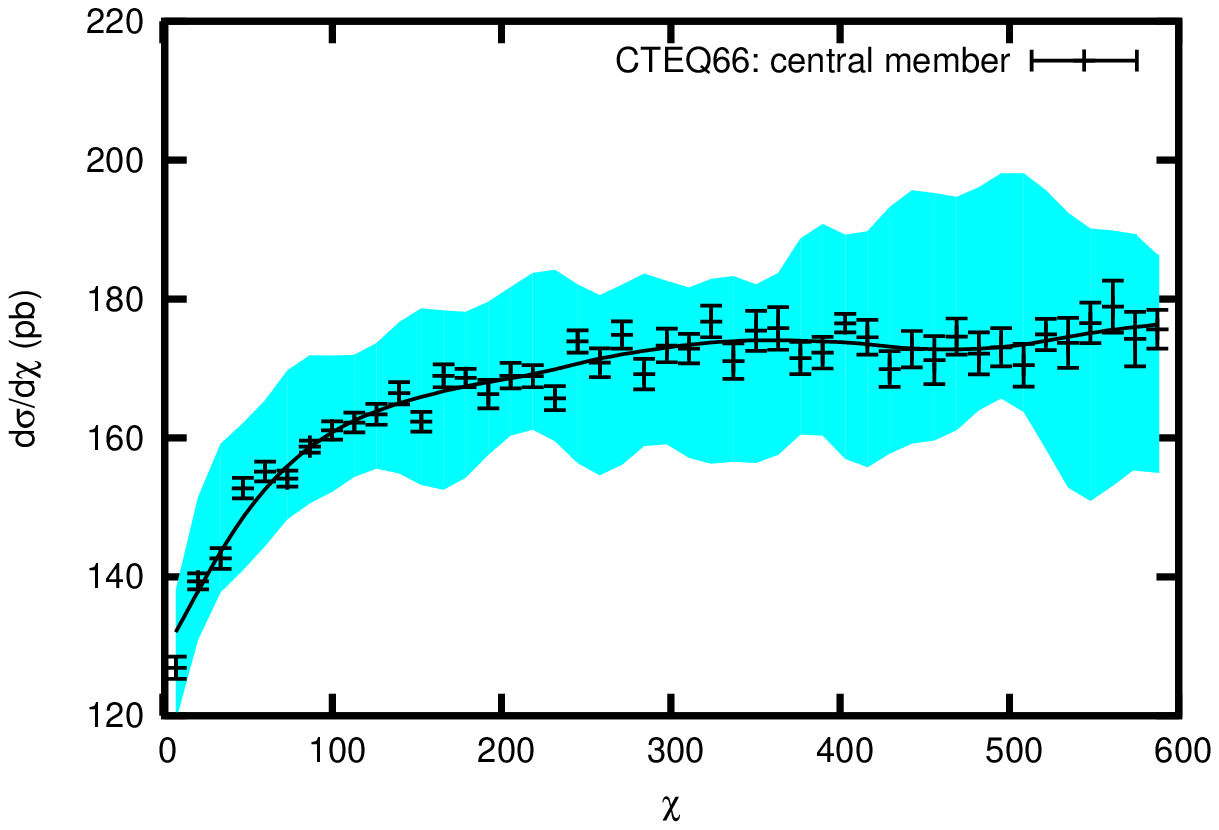}
\caption{Central member of the CTEQ66 PDF, together with its uncertainty band for the mass bin $1 < M_{jj} < 2$ TeV. The calculations are done with JETRAD and an inclusive $k_T$ algorithm with R = 1.0.}
\label{CTEQ66}
\end{center}
\end{figure*}

\begin{figure*}
\begin{center}
\includegraphics[angle=-90,width=0.8\textwidth]{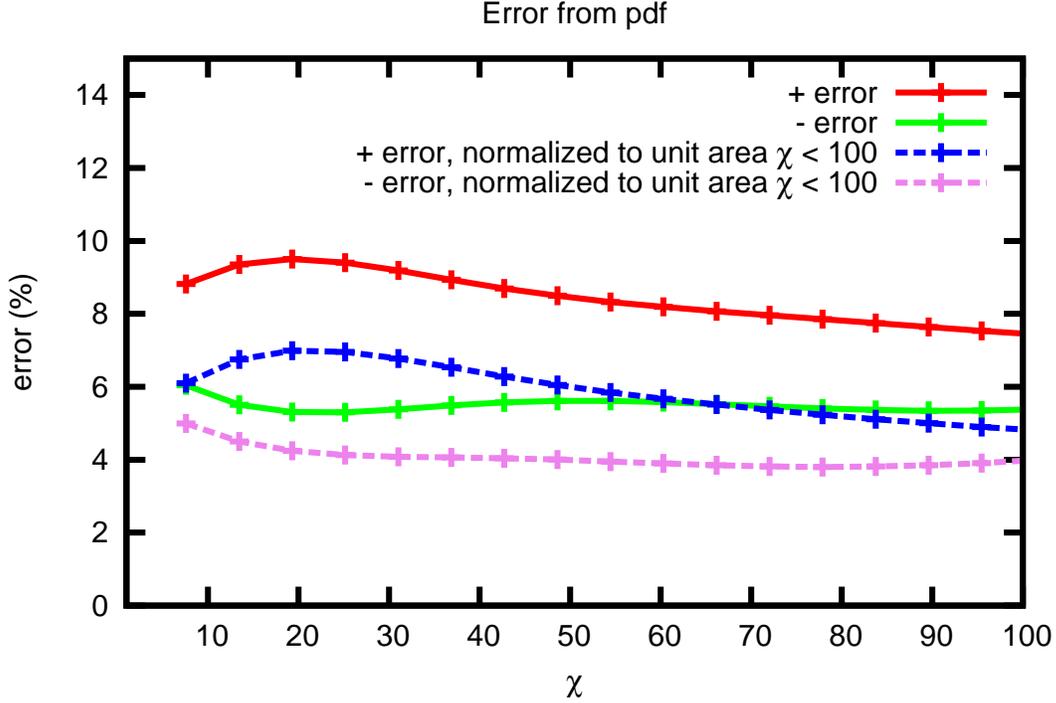}
\caption{Uncertainty on the angular distribution calculated with the central member of the CTEQ66 PDF, in the mass bin $1 < M_{jj} < 2$ TeV coming from the intrinsic uncertainty of the PDF. The $+$ error leads to an increase of the cross section, while the $-$ error leads to a decrease (see equation (\ref{uncertaintyEq}) in the text). The uncertainties are shown on the curves both with and without normalization to unit area $1 < \chi < 100$.}
\label{cteq66_error}
\end{center}
\end{figure*}

To further examine the uncertainties coming from PDFs, we have calculated the angular distributions for all 44 error members of the CTEQ66 PDF and applied the Master Equation suggested in \cite{LHCPrimer}  to deduct a positive and negative uncertainty on a quantity $X$:
\begin{gather}
\Delta X^+_{\mathrm{max}} = \sqrt{ \sum_{i=1}^N [\max(X^+_i-X_0,X^-_i-X_0,0 ) ]^ 2}\\
\Delta X^+_{\mathrm{min}} = \sqrt{ \sum_{i=1}^N [\max(X_0-X^+_i,X_0-X^-_i,0 ) ]^ 2}
\label{uncertaintyEq}
\end{gather}
$\Delta X^+$ adds in quadrature the PDF error contributions that lead to an increase in the observable $X$, and $\Delta X^-$ the PDF error contributions that lead to a decrease.
Using this formula, the error on the dijet angular distribution calculated with the central PDF member, in the mass bin $1 < M_{jj} < 2$ TeV is plotted in figure \ref{CTEQ66} as a blue band around the central member.
The uncertainties given in terms of percentage are shown in figure \ref{cteq66_error}.

\begin{figure*}
\begin{center}
\includegraphics[width=1.0\textwidth]{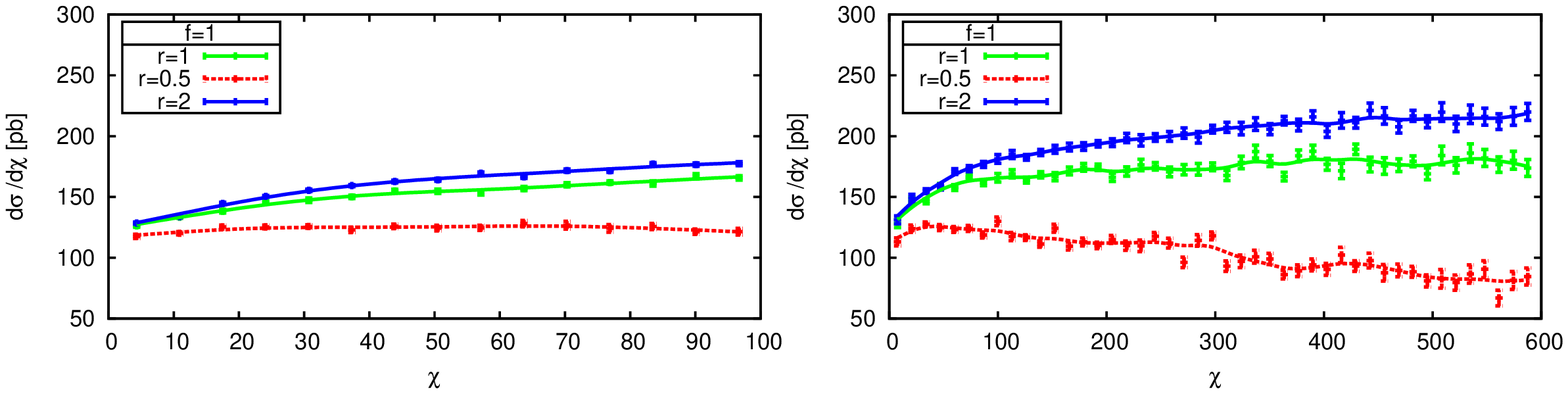}
\caption{Influence of the choice of $\mu_{R}$ for the mass bin $1 < M_{jj} < 2$ TeV. The variables $r$ and $f$ denote the fraction of the transverse momentum of the highest jet at which respectively the renormalization and factorization scale are evaluated. Left: $\chi<100$, right: $\chi<600$. The calculations are done with JETRAD and an inclusive $k_T$ algorithm with R = 1.0.}
\label{norm_fac1}
\end{center}
\end{figure*}

\begin{figure*}
\begin{center}
\includegraphics[width=1.0\textwidth]{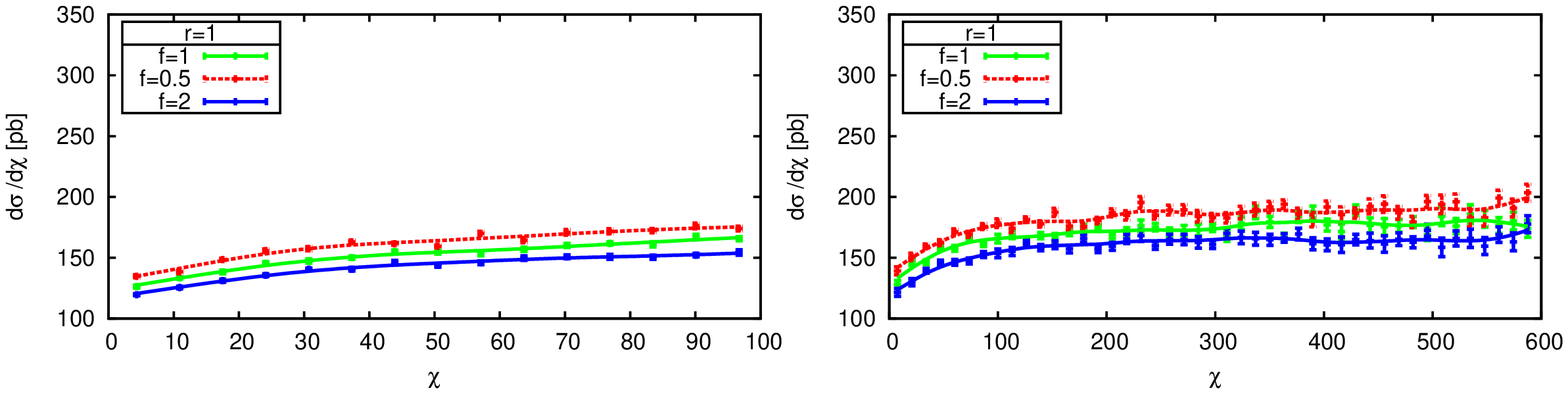}
\includegraphics[width=1.0\textwidth]{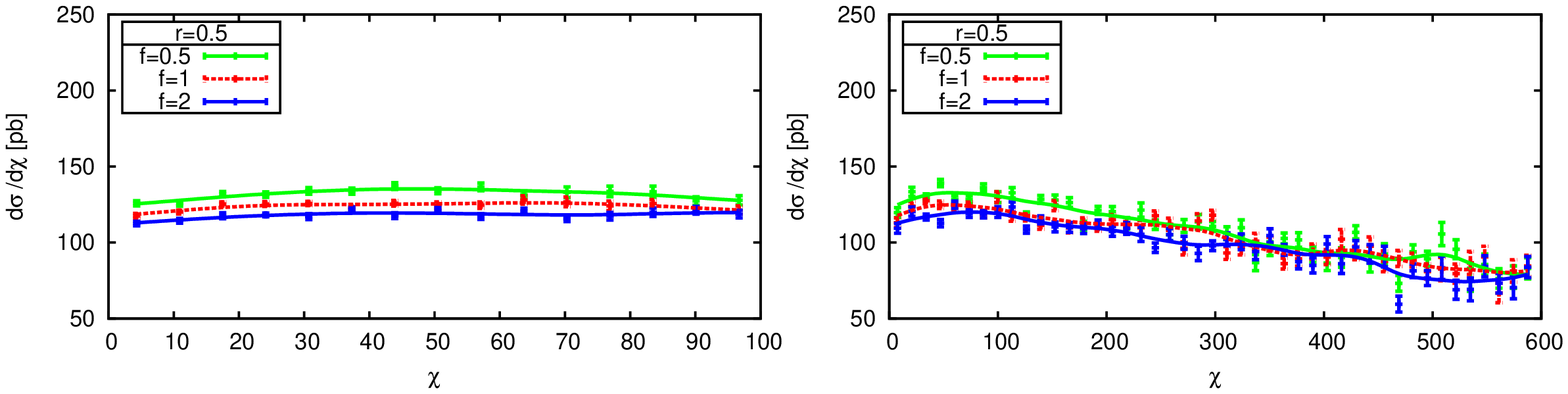}
\includegraphics[width=1.0\textwidth]{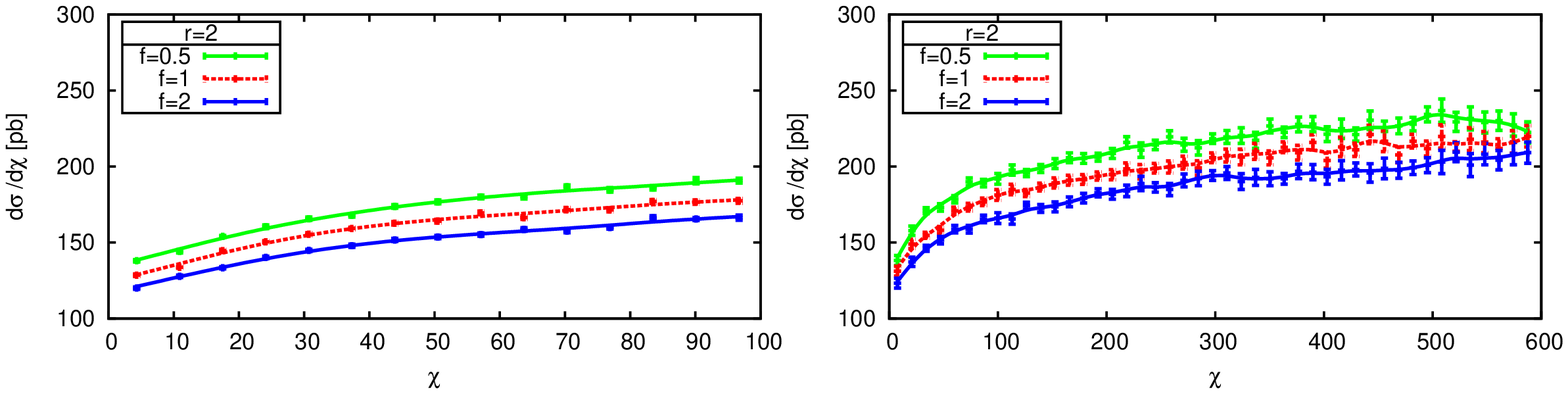}
\caption{Influence of the choice of $\mu_{F}$ for the mass bin $1 < M_{jj} < 2$ TeV for different values of $\mu_{R}$. The variables $r$ and $f$ denote the fraction of the transverse momentum of the highest jet at which respectively the renormalization and factorization scale are evaluated. Left: $\chi<100$, right: $\chi<600$. The calculations are done with JETRAD and an inclusive $k_T$ algorithm with R = 1.0.}
\label{norm_fac2}
\end{center}
\end{figure*}

The choice of $\mu_R$ and $\mu_F$ will also influence the distributions. We have studied this by letting $\mu_R$ and $\mu_F$ vary independently between 0.5, 1 and 2 times the transverse momentum of the hardest jet, resulting in 9 different distributions in total. Figures \ref{norm_fac1} and \ref{norm_fac2} summarize the results for the mass bin $1 < M_{jj} < 2$ TeV. The other mass bins have similar results. For all these figures, the plots at the left show the distributions for $\chi<100$, while the plots at the right go up to $\chi=600$. The variables $r$ and $f$ used in the figures, denote the fraction of the transverse momentum of the highest jet at which respectively the renormalization and factorization scale are evaluated.  
In figure  \ref{norm_fac1}, $\mu_F$ is kept constant, while $\mu_R$ is varied, which influences both the shape and the normalization of the distributions. The effect is small at low $\chi$ but increases drastically with increasing $\chi$ values. The plots in figure \ref{norm_fac2} all have fixed $\mu_R$ and varying $\mu_F$, which causes a change rather in normalization and not so much in shape.

\begin{figure*}
\begin{center}
\includegraphics[angle=-90,width=0.7\textwidth]{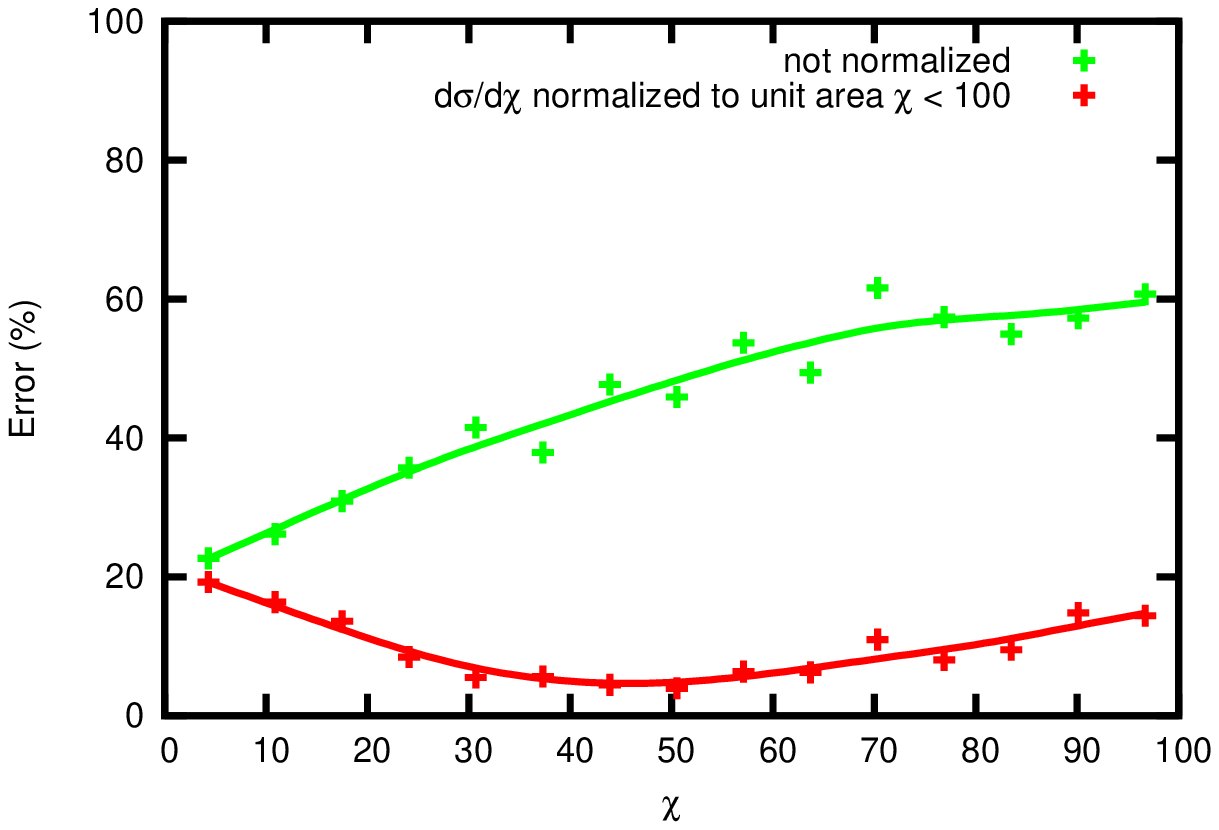}
\caption{Systematic uncertainty coming from the choice of $\mu_{R,F}$ for the mass bin $1 < M_{jj} < 2$ TeV, both on normalized and not normalized distributions.}
\label{error_mur_muf}
\end{center}
\end{figure*}

The difference between these distributions is an estimate of the uncertainty coming from the choice of $\mu_R$ and $\mu_F$. In figure \ref{error_mur_muf} we show the uncertainty on the distributions both with and without normalization to unit area $1 < \chi < 100$. Normalizing the distributions reduces the error drastically; averaged over the whole $\chi$ range, the uncertainty on the normalized distribution is 9$\%$, and is never exceeding 20$\%$. 

\begin{figure*}
\begin{center}
\includegraphics[width=1.0\textwidth]{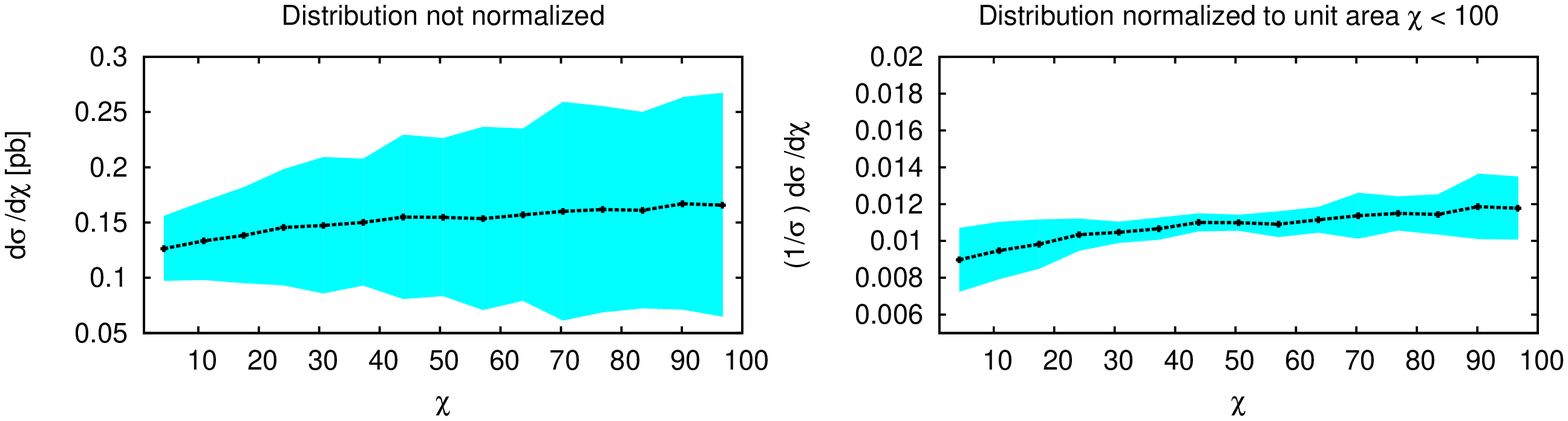}
\caption{Black line: calculation done with nominal central CTEQ66 member and $\mu_R = mu_F = p_T$ of the highest jet, both with (plot right) and without (plot left) normalization to unit area $1 < \chi < 100$. Blue band: error band from combining the uncertainties coming from the choice of renormalization and factorization scale (figure \ref{error_mur_muf}), together with the intrinsic uncertainty from the CTEQ66 PDF (figure \ref{cteq66_error}). }
\label{comErrors}
\end{center}
\end{figure*}
Figure \ref{comErrors} shows the combination in quadrature of the uncertainties coming from the choice of renormalization and factorization scale (figure \ref{error_mur_muf}), together with the intrinsic uncertainty from the CTEQ66 PDF (figure \ref{cteq66_error}), drawn as an error band around the calculation done with nominal values (central CTEQ66 member and $\mu_R = \mu_F = p_T$ of the highest jet). Both the distributions with (plot right) and without (plot left) normalization to unit area $1 < \chi < 100$ are shown.
In both plots, the error band is dominated by the error coming from the choice of the scales, more precisely, the renormalization scale introduces the major uncertainty. When normalized to unit area $1< \chi < 100$, the combined error does not exceed 20$\%$ over the whole $\chi$ range.  
\par
\begin{figure*}
\begin{center}
\includegraphics[angle=-90,width=0.8\textwidth]{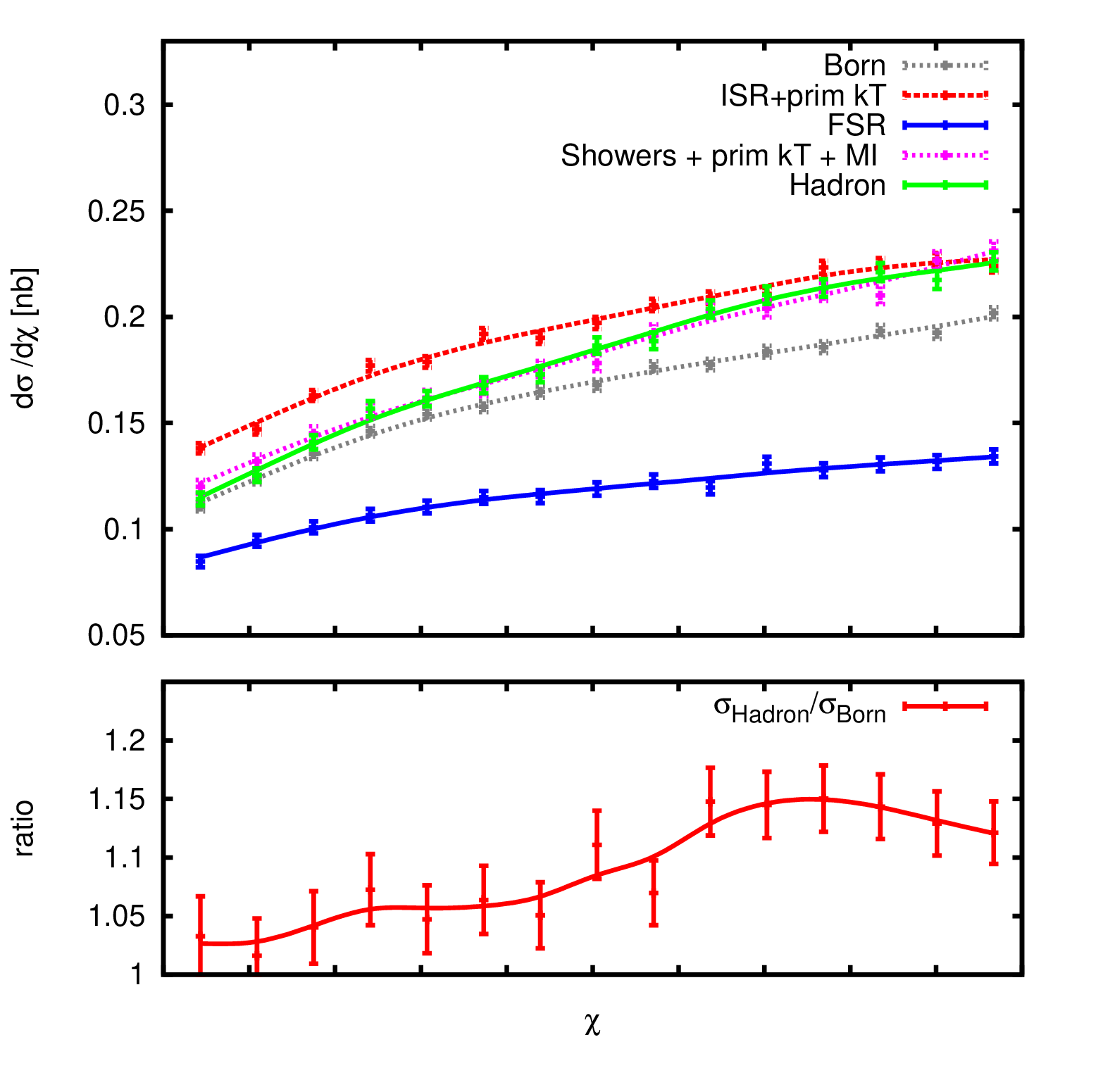}
\caption{Including primordial $k_T$, initial and final state radiation, multiple interactions and hadronization in a LO calculation. The calculations are done using PYTHIA and the PYCELL cluster routine. }
\label{hadrons}
\end{center}
\end{figure*}
Above calculations have been done at the parton level of the hard interaction, without showering, multiple interactions or hadronization. In figure \ref{hadrons} we show what happens if we turn on these processes. In the top figure, we have plotted angular distributions calculated with PYTHIA 6.410 \cite{pythia}. The dashed gray curve is a calculation at the Born level, without switching on any of the processes that were previously mentioned. Initial state radiation (ISR) and primordial $k_T$ (prim kT) have been turned on for the dashed red curve and the full blue curve is a calculation with final state radiation (FSR) included (but initial state radiation and primordial $k_T$ turned off). As can be observed in the figure, final state radiation causes losses out of the jet cone. The dashed magenta curve covers both initial and final state showers, primordial $k_T$ and multiple interactions. Finally, the full green curve includes all these processes and hadronization.
The calculations are done with the PYCELL cluster routine from the PYTHIA library and the PYTHIA default settings for initial and final state radiation, multiple interactions and hadronization. For this particular choice of jet algorithm and settings, the difference between the calculation at the Born level and the one at the hadron level is rather small, as can be seen from their ratio, shown in the bottom plot of figure \ref{hadrons}. The corrections may vary for different jet algorithms and PYTHIA settings. 

\section{Gravitational scattering and black hole formation in large extra dimensions}
\label{QG}
An attempt to resolve the hierarchy problem in particle physics, i.e.~why the electroweak scale is so many orders of magnitude lower than the observed Planck scale, was done by Arkani-Hamed, Dimopoulos and Dvali in the late nineties \cite{arkanihamed-1998-429,arkanihamed-1999-59}. Their so-called ADD model assumes the existence of large extra spatial dimensions in which gravity is allowed to propagate, while the SM fields are confined to a four-dimensional membrane. 
The fact that the extra dimensions have not been observed yet can be explained by assuming that they are compactified with a (common) compactification radius $R$. This causes the fundamental Planck scale $M_{P}$ to be much smaller than the observed 4-dimensional one, $M_{P4}$, i.e.~for \mbox{n extra}  dimensions the relation $M^2_{P4} \sim M^{n+2}_{P}R^n $ holds. Experiments conducted to test the gravitational inverse-square law have found that an extra dimension must have a size $R \leq 44$ $\mu m$ \cite{kapner-2007-98}. 

For a fundamental Planck scale of around 1 TeV, the ADD model predicts the production of extra dimensional black holes at the LHC. Besides black holes, also processes involving the exchange of virtual Kaluza-Klein (KK) modes, with gravitational scattering of hard partons as dominant process, will be present.

Effects from Black Hole production and gravitational scattering on standard jet observables such as the $p_T$-spectrum and the dijet invariant mass spectrum were investigated in \cite{lonnblad-2006-0610}. The study was based on an effective field description of gravitational scattering done in \cite{han-1999-59,giudice-1999-544,sjodahl-2006}. This section will continue this study by elaborating on dijet angular distributions. First we will give a brief summary of the theory, then we will show results.

\subsection{Gravitational scattering}
\label{GS}

In a gravitational scattering event, quantized KK modes will occur as intermediate states and, in order to calculate the scattering amplitude, the sum over all modes has to be made. This sum (which can be turned into an integral) diverges, but can be rendered finite by introducing a physical cut-off of the KK tower to be summed over. A cut-off implied from a narrow (small compared to the compactification radius $R$) width of the four-dimensional membrane, assuming a Gaussian extension of the standard model field densities into the bulk, was proposed in \cite{sjodahl-2006} and a calculation of the $t$-channel amplitude was made. The finite width suppresses the coupling to high-mass KK modes with a factor $e^{-m^2/M_s^2}$, $m$ being the mass of the mode and $1/M_s$ the width of the four-dimensional membrane, and an effective propagator can be derived. The scattering process now depends on three energy scales, namely the collision energy $\sqrt{\hat{s}}$, the fundamental Planck scale $M_{P}$ and the inverse of the membrane width $M_s$.

For $\sqrt{\hat{s}} \gg M_s$, forward scattering via the $t$-channel dominates and an all-order eikonal calculation was done to ensure unitarity. Contributions from multi-loop ladder diagrams exponentiate  and the all order eikonal amplitude is given by 
\begin{equation}
  A_{\mathrm{eik}}(k^2)=-2i\hat{s} \int d^2\bar{b}_{\perp}e^{i\bar{k}_{\perp}\cdot\bar{b}_{\perp}}(e^{i\chi(b)}-1),
\end{equation}
with $b$ the impact parameter and $\chi(b)$ the eikonal scattering phase:
\begin{equation}
\chi(b) = \frac{1}{2\hat{s}}\int \frac{d^2\bar{k}_{\perp}}{(2\pi)^2}e^{-i\bar{k}_{\perp}\cdot\bar{b}_{\perp}}A_{\mathrm{Born}}(-\bar{k}^2_{\perp} ),
\label{eikonal}
\end{equation}
The Born term used in equation (\ref{eikonal}) is dominated by small values of $ (\hat{t}/\hat{s})$, i.e.~small angle scattering, and is ---neglecting spins of colliding partons--- given by:
\begin{equation}
A_{\mathrm{Born}}(\hat{t}) =  \frac{1}{2^{n-3}\pi^{\frac{n-2}{2}}} \Big{(} \frac{Ms}{M_P} \Big{)}^n \frac{\hat{s}^2}{M_P^2\cdot \hat{t}}
\end{equation}

On the other hand, when the exchanged momentum is small compared to $M_s$, i.e.~for $\sqrt{\hat{s}}<M_s$, $s$-,$t$- and $u$-channels are equally efficient and scattering is almost isotropic. KK propagators in the $t$-channel can be replaced by vertex factors so that the exchange of KK modes corresponds to a contact interaction. A geometric series of ladder diagrams is obtained and summed over so that unitarity is constrained:
 \begin{equation}
 A_{\mathrm{ladders} } = \frac{ A_{\rm{Born} } }{ 1-A_{\rm{Born} }X},
\end{equation}
with $X \approx  \frac{1}{32 \pi^2} ( \ln\frac{M_s^2}{\hat{s}/4} + i\pi) $.
The Born term ---neglecting spins of the colliding partons--- can be written in terms of an effective Planck mass  $M_{\mathrm{eff}}$:
\begin{equation}
A_{\mathrm{Born}}=-\frac{\hat{s}^2}{M_{\mathrm{eff}}^4}
\end{equation}
with:
 \begin{equation}
M_{\mathrm{eff}}=\frac{1}{2}(\frac{(n-2)2^n\pi^{\frac{n-2}{2}}M^{n+2}_{P}}{M^{n-2}_s})^{\frac{1}{4}}
\label{EffectiveMass}
\end{equation}
The $s$-channel ladder diagrams (rotated by $\pi/2$) are unitarized in a similar way. KK modes can also be produced on shell in this channel, but with a lifetime too long to detect. 

The relevance of the $u$-channel contribution is suppressed for proton-proton collisions because of the low probability of a collision between two partons with identical flavor, spin and color. 

Note that gravitational scattering will become weaker for smaller values of $M_s$ (i.e.~wider membranes), because the KK-modes with a mass above $M_s$ are suppressed.

\subsection{Black holes}
\label{BH}
The Schwarzschild radius, introduced by Schwarzschild for static non-spinning massive objects, can be generalized to 4+n dimensions \cite{Myers1986304}:
\begin{equation}
r_s \propto \frac{1}{M_P}(\frac{M_{\mathrm{BH}}}{M_P})^{\frac{1}{n+1}}
\label{rsch}
\end{equation}
In hadron collisions, black holes can be formed when the interacting partons come closer than twice $r_s$. This means that in the rest frame of the incoming partons their longitudinal wavelength $\lambda_l \propto 2/\sqrt{\hat{s}}$ and transverse wavelength $\lambda_{\bot} \propto 1/p_T$ need to be smaller than $r_s$. This implies a minimum on the black hole mass, which lies in the neighborhood of the Planck scale \cite{1126-6708-2005-09-019}:
\begin{equation}
M_{\mathrm{min,1}}= M_P (\frac{ (2\sqrt\pi)^{n+1}(n+2)}{8 \Gamma(\frac{n+3}{2})})^{\frac{1}{n+2}}
\label{BH_mass}
\end{equation}

Another limit on the black hole's mass comes from the existence of the finite width of the membrane, because, in the approximation of a narrow width, the membrane cannot be more extended than the black hole itself. This means that the Schwarzschild radius $r_s$ should not be smaller than $1/M_s$, which is only true for masses above $M_{\mathrm{min,2}}$:
\begin{equation}
M_{\mathrm{min,2}}= \frac{M^{n+2}_{P}(2+n)\pi^{\frac{n+1}{2}}}{8\Gamma (\frac{3+n}{2}) M^{n+1}_s}
\label{BH_mass1}
\end{equation}
The maximum of (\ref{BH_mass1}) and (\ref{BH_mass}) is the minimum black hole mass possibly created. For n=4 extra dimensions and for $M_s/M_P > 1.12$ ($M_s/M_P > 1.05$ for n=6), the minimum black hole mass equals (\ref{BH_mass}) and depends on $M_P$ only. In the complementary region the minimum black hole mass is described by (\ref{BH_mass1}) and goes as $M_P/ (M_s/M_P)^5$ ($M_P/ (M_s/M_P)^7$ for n=6). In the latter case, small values of $M_s$, i.e.~larger values of the membrane width, will prevent black holes from being created.

\subsection{Results}
\label{Results}

Gravitational scattering was implemented in the PYTHIA 6.410 event generator and combined with the CHARYBDIS \cite{harris-2003-0308} black hole generator, in a similar way as in \cite{lonnblad-2006-0610}. The calculations have been done using an inclusive $k_T$ algorithm with separation parameter 1.0. We have used the k-factor which was derived in the previous section (see figure \ref{Mjj}) to scale the PYTHIA QCD distributions up to NLO.
 
As discussed in the previous sections, the model parameters determining the phenomenology are $M_{P}$, n and $M_s$. An equivalent set of parameters is $M_{\mathrm{eff}}$ (equation (\ref{EffectiveMass})), $M_s/M_{P}$ and n, and we will work with the latter one. Table \ref{t2} shows a few combinations. All parameter choices in table \ref{t2} have been tested at a Tevatron energy of $\sqrt{s}=1.8$ TeV for the mass bin  635 GeV $< M_{jj}$ and for  $\chi \leq 20$, and turned out to be consistent with the dijet angular distribution measurements done by the CDF \cite{PhysRevLett.77.5336} and D0 collaboration \cite{PhysRevLett.80.666}.
\begin{table*}
\begin{center}
\caption{Different parameter sets}
\begin{tabular}{c c c c c c}
\hline  \hline
 Name & $M_{\mathrm{eff}}$ (TeV)	& n  & $M_s/M_{P}$	& $M_{P}$ (TeV) &  $M_s$ (TeV) \\
\hline
C1 & 1.0 & 6 & 1.0 & 0.282 & 0.282 \\
C2 & 1.0 & 6 & 2.0 & 0.564 & 1.128 \\
C3 & 1.0 & 6 & 4.0 & 1.128 & 4.513 \\
C4 & 0.5 & 6 & 8.0 & 1.128 & 9.027 \\
C5 & 1.0 & 6 & 8.0 & 2.257 & 18.05 \\
C6 & 1.0 & 4 & 4.0 & 1.263 & 5.053 \\
 \hline
\end{tabular}
\label{t2}
\end{center}
\end{table*}

\begin{figure*}
\begin{center}
\includegraphics[angle=-90,width=0.48\textwidth]{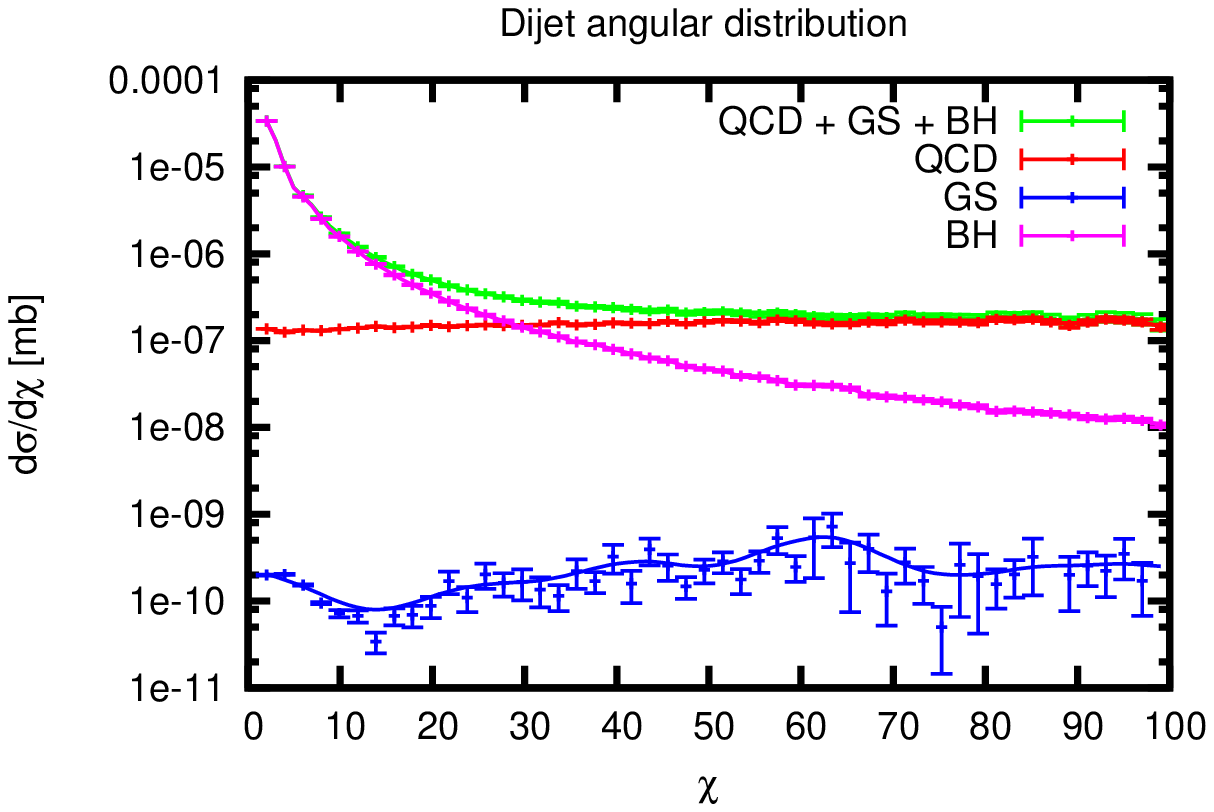}
\hfill
\includegraphics[angle=-90,width=0.48\textwidth]{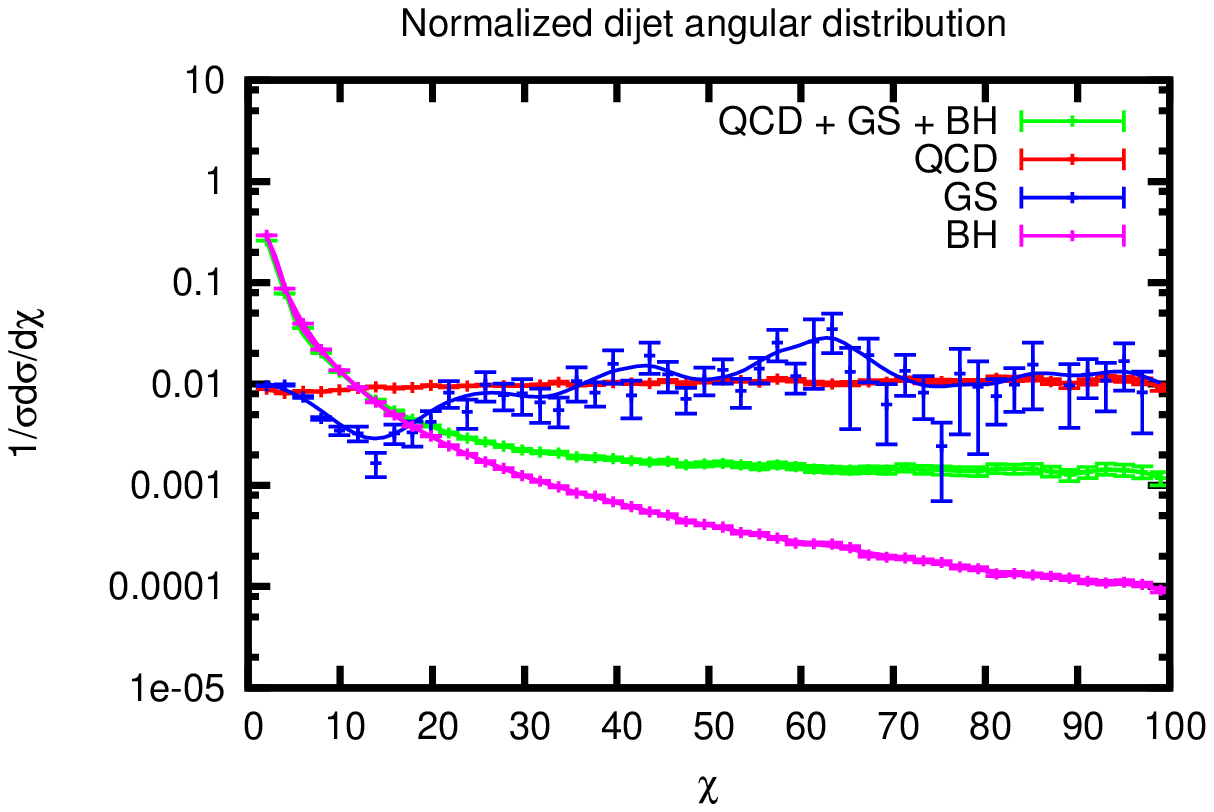}
\caption{Angular distributions for C1 (see table \ref{t2}) for the mass bin $1 < M_{jj} < 2$ TeV. Left: cross section in mb, right: cross section normalized to unit area $1< \chi < 100$. }
\label{C1_12}
\end{center}
\end{figure*}

\begin{figure*}
\begin{center}
\includegraphics[angle=-90,width=0.48\textwidth]{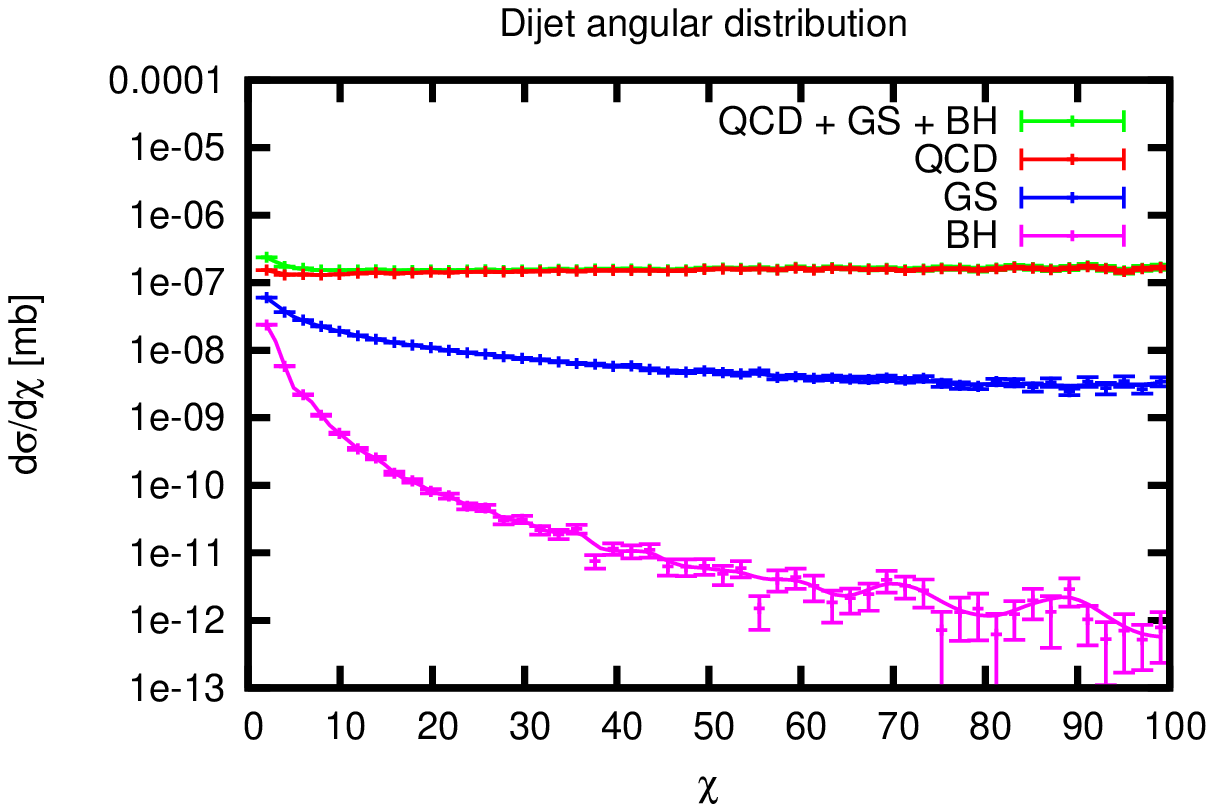}
\hfill
\includegraphics[angle=-90,width=0.48\textwidth]{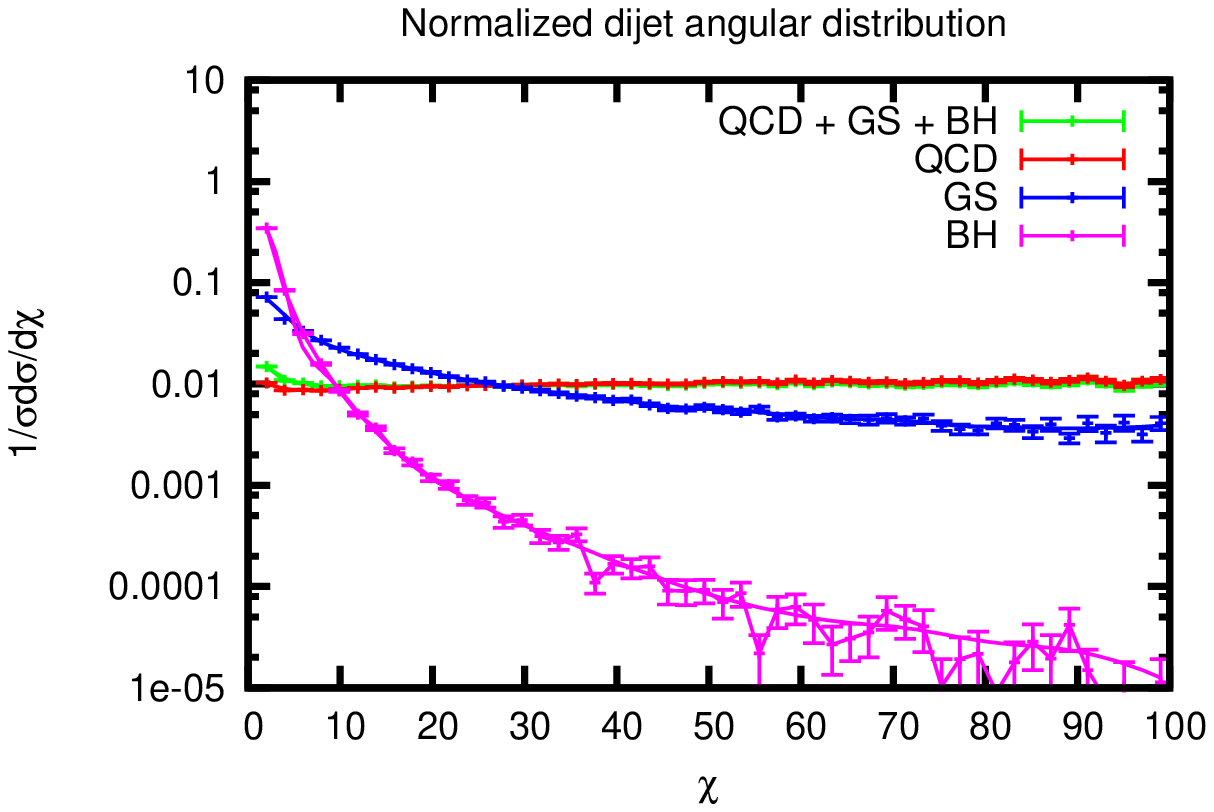}
\caption{Angular distributions for C3 (see table \ref{t2}) for the mass bin $1 < M_{jj} < 2$ TeV. Left: cross section in mb, right: cross section normalized to unit area $1< \chi < 100$.}
\label{C3_12}
\end{center}
\end{figure*}

\begin{figure*}[ht!]
\begin{center}
\includegraphics[angle=-90,width=0.48\textwidth]{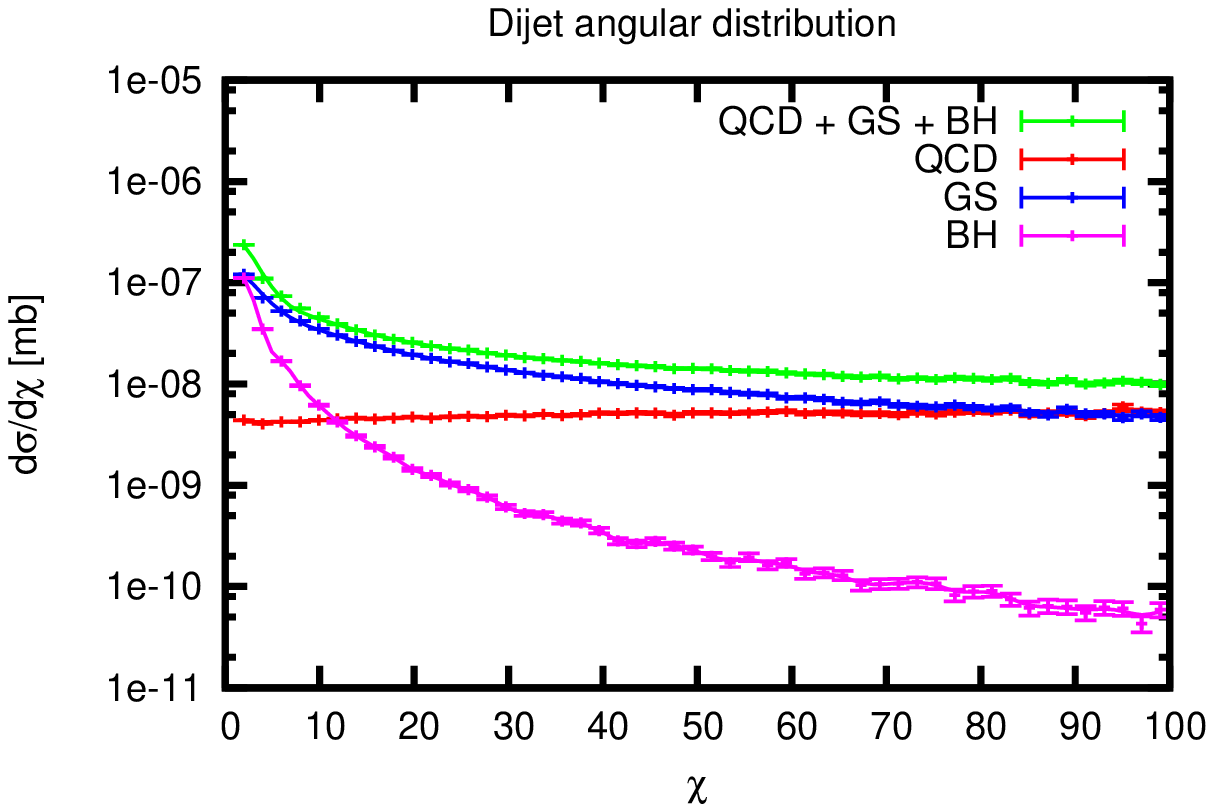}
\hfill
\includegraphics[angle=-90,width=0.48\textwidth]{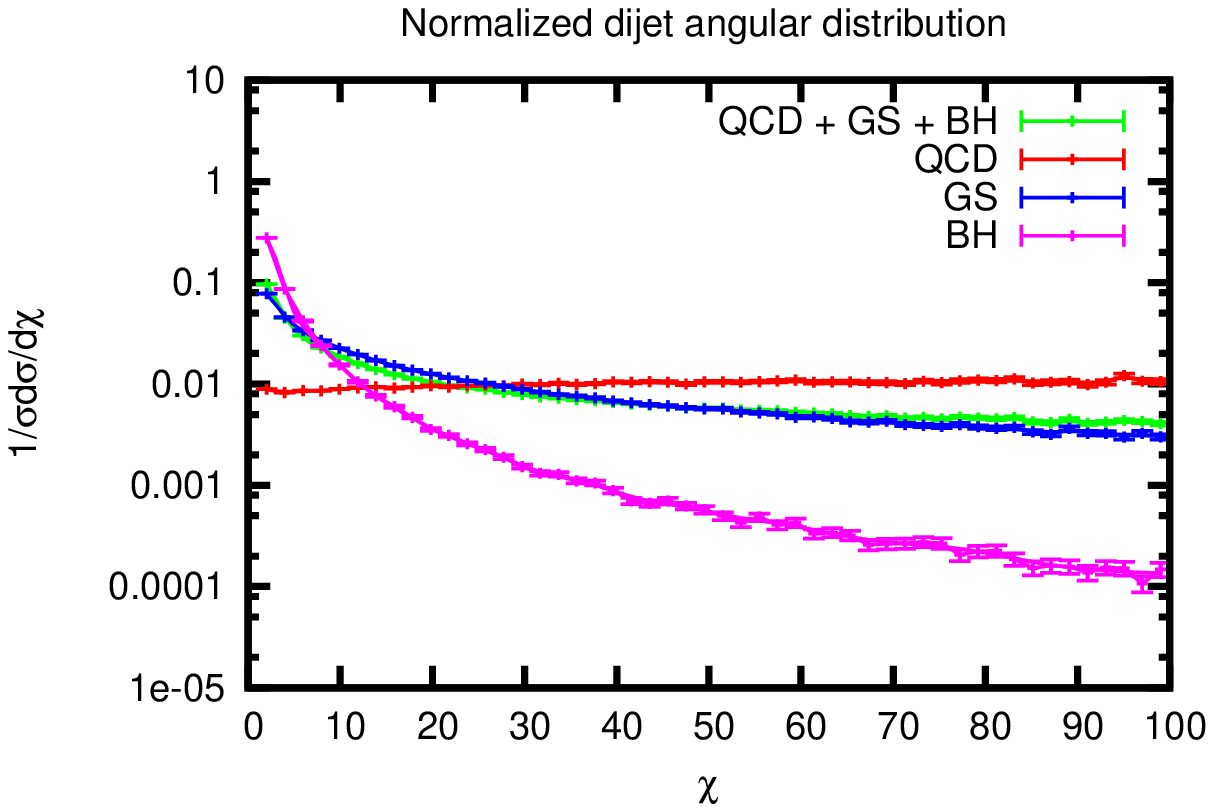}
\caption{Angular distributions for C3 (see table \ref{t2}) for the mass bin $2 < M_{jj} < 3$ TeV. Left: cross section in mb, right: cross section normalized to unit area $1 < \chi < 100$.}
\label{C3_23}
\end{center}
\end{figure*}

\begin{figure*}
\begin{center}
\includegraphics[angle=-90,width=0.48\textwidth]{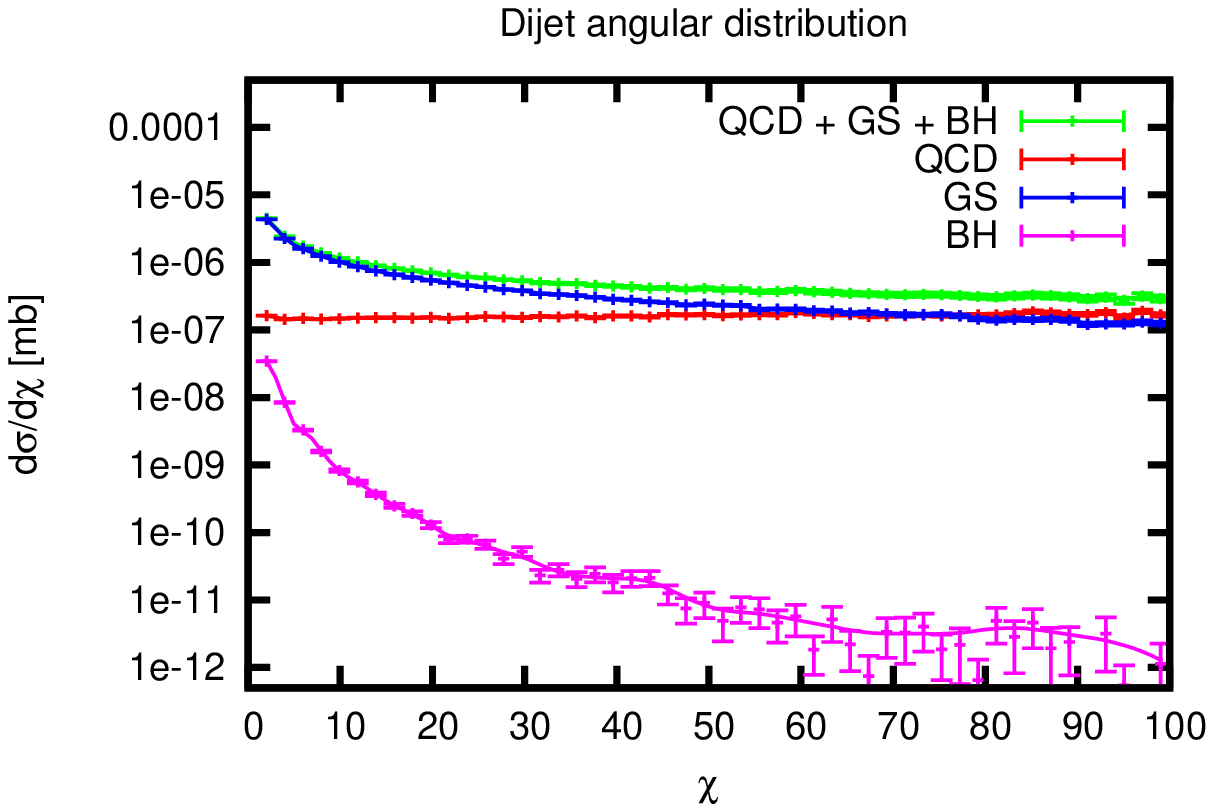}
\includegraphics[angle=-90,width=0.48\textwidth]{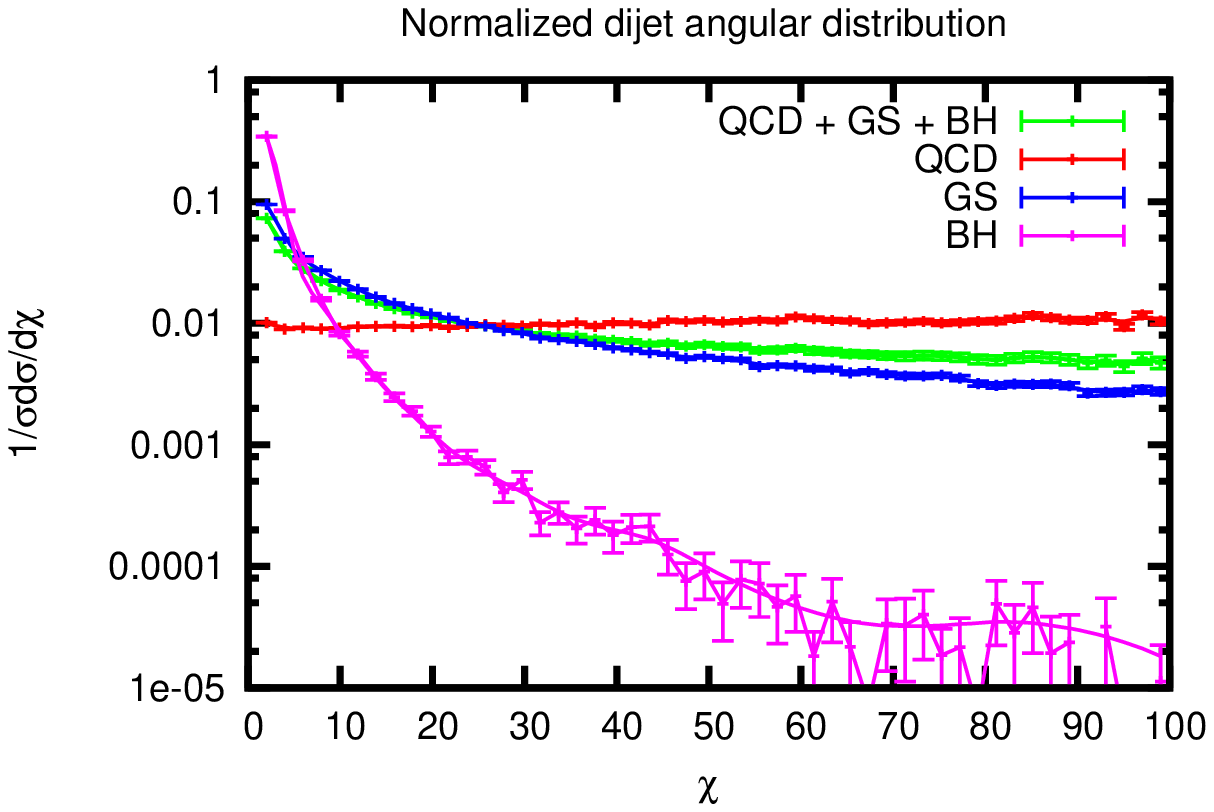}
\caption{Angular distributions for C4 (see table \ref{t2}) for the mass bin $1 < M_{jj} < 2$ TeV. Left: cross section in mb, right: cross section normalized to unit area $1< \chi < 100$.}
\label{C4_12}
\end{center}
\end{figure*}

\begin{figure*}
\begin{center}
\includegraphics[angle=-90,width=0.48\textwidth]{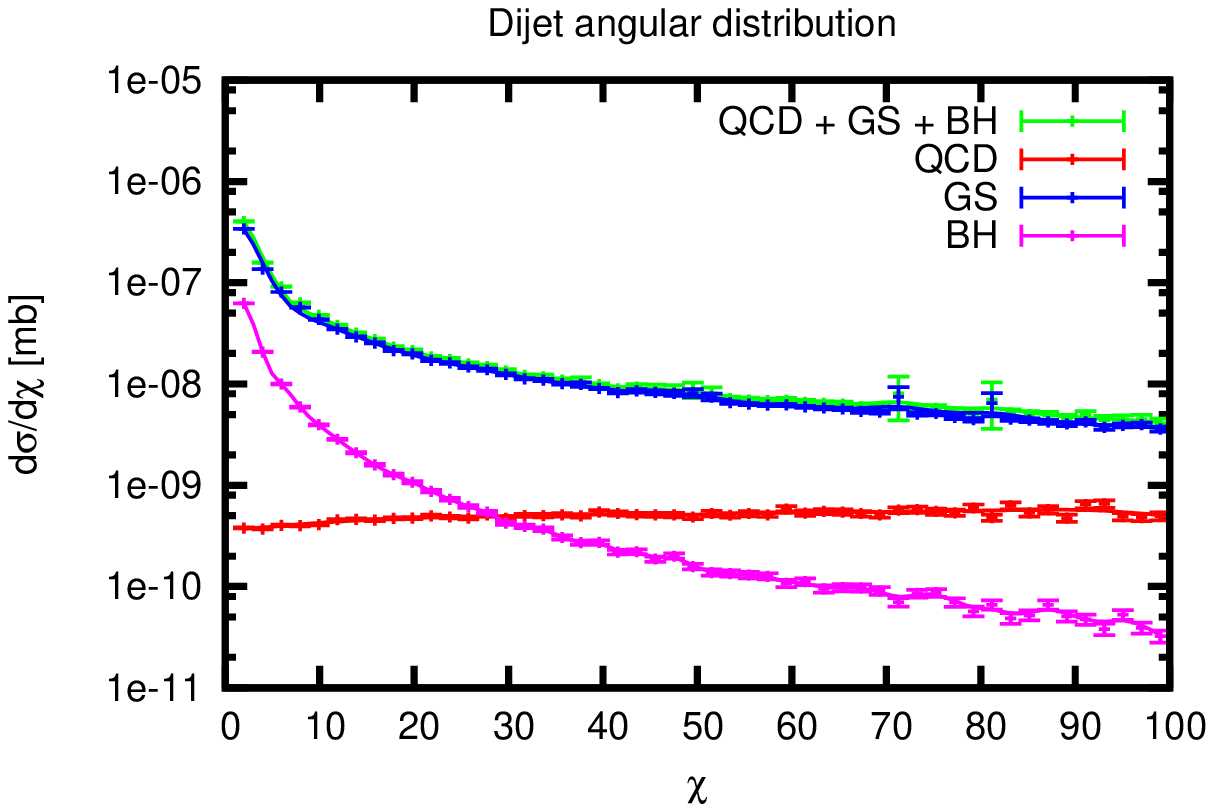}
\hfill
\includegraphics[angle=-90,width=0.48\textwidth]{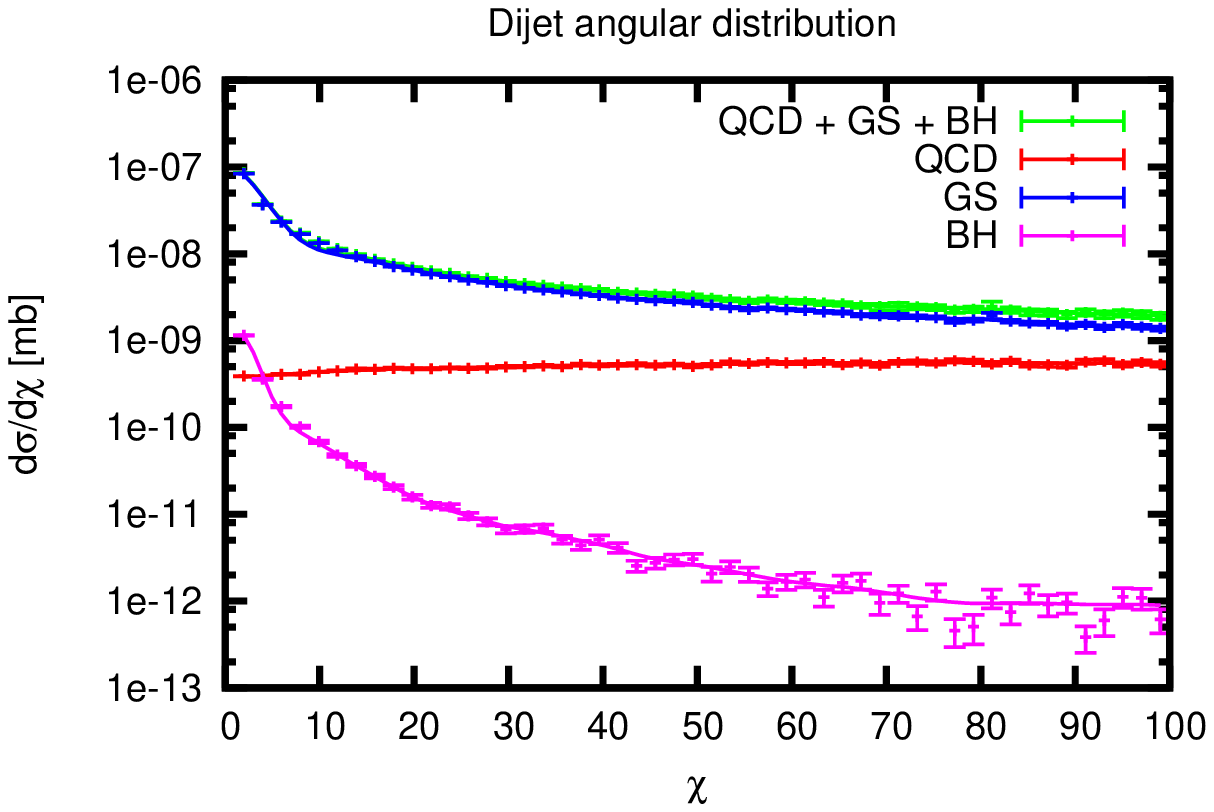}
\caption{Angular distributions (in mb) in the mass bin 3 TeV $< M_{jj}$ for C4 (left) and C5 (right).}
\label{C45_314}
\end{center}
\end{figure*}

\begin{figure*}
\begin{center}
\includegraphics[angle=-90,width=0.48\textwidth]{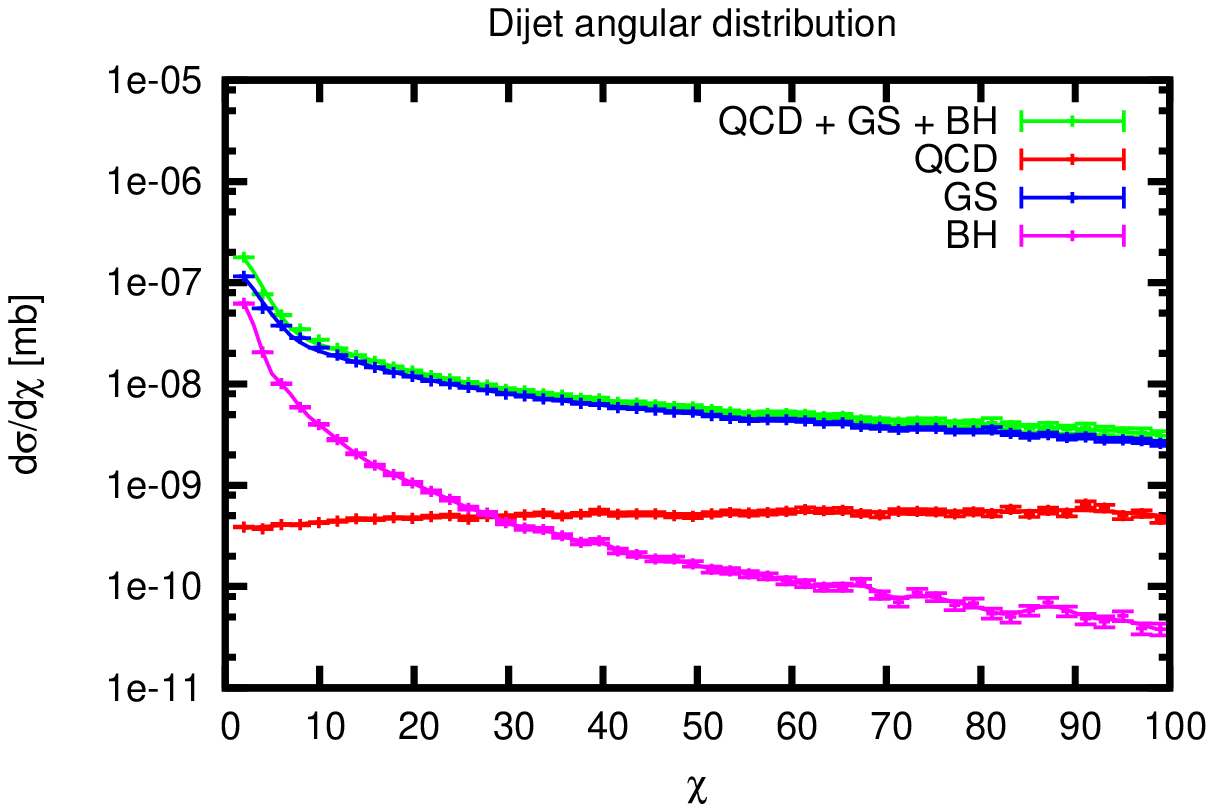}
\hfill
\includegraphics[angle=-90,width=0.48\textwidth]{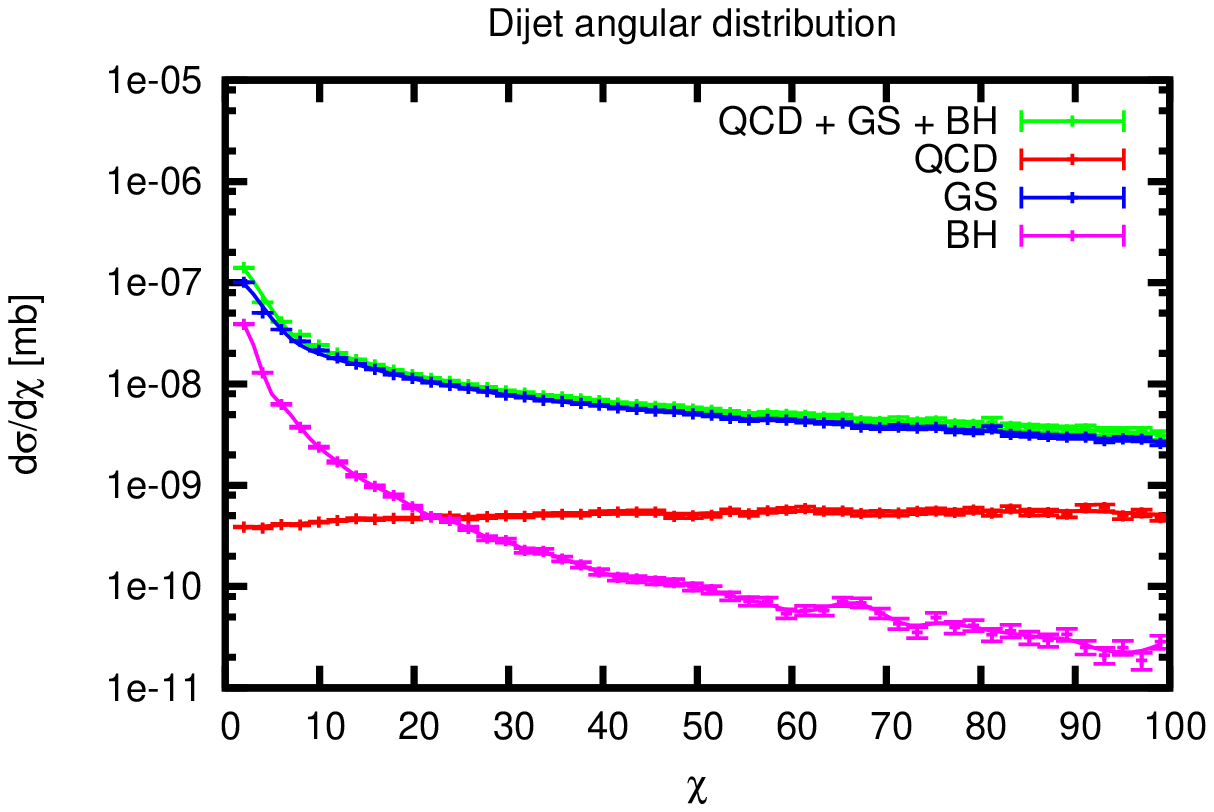}
\caption{Angular distributions (in mb) in the mass bin 3 TeV $< M_{jj}$ for C3 (left) and C6 (right).}
\label{C36_314}
\end{center}
\end{figure*}

To get a feeling of the impact of the parameters, we have performed a few runs with PYTHIA with parton showers, multiple interactions and hadronization turned off. Figures \ref{C1_12} to \ref{C36_314} show the dijet angular distributions, with and without normalizing to unit area $1 < \chi < 100$, for some of the parameter sets defined in table \ref{t2}. Previous experiments have shown that normalizing the distributions reduces the experimental error \cite{PhysRevLett.77.5336,PhysRevLett.80.666}. Furthermore, in the previous section we have demonstrated that the theoretical error on normalized distributions is smaller as well (figure \ref{comErrors}). We have plotted the different contributions  ---i.e.~gravitational scattering (GS), black holes (BH) and QCD--- separately, as well as their sum. As with most new physics, the biggest effects from gravitational scattering and black hole formation show up at low $\chi$ (high $p_T$) values. How much each process contributes to the total cross section, depends on the parameter settings and the mass bin. We will make a few quantitative observations.
\par
Increasing $M_s/M_{P}$ while keeping $M_{\mathrm{eff}}$ fixed, will result in an increase of gravitational scattering and a decrease of black hole formation. The effect is clearly visible when set C1 (figure \ref{C1_12}) is compared with set C3 (figure \ref{C3_12}) for the mass bin $1 < M_{jj} < 2$ TeV. For C1, the total cross section at low  $\chi$ values, is dominated by black holes, while gravitational scattering is of no importance. But this changes drastically for C3; black holes have almost completely disappeared for C3, but the gravitational scattering cross section has increased by two orders of magnitude. QCD is still dominating in this mass bin, but gravitational scattering starts to dominate the cross section for mass bins $\geq 2$ TeV, as can be seen in figure \ref{C3_23}.  
\par
Compared to C3, C4 has the same fundamental Planck scale and number of extra dimensions (1 TeV and 6, respectively), but a different width of the membrane (9 TeV vs 4.5 TeV), and this causes effects from gravitational scattering to set in at lower mass values; from figure \ref{C4_12}, it is observed that gravitational scattering dominates the cross section from 1 TeV onwards, which is not the case for C3 (figure \ref{C3_12}). 
\par 
Figure \ref{C45_314} examines what happens if we double the effective and fundamental Planck scale; we have plotted the angular distribution in the mass bin 3 TeV $< M_{jj}$ TeV for C4 (figure \ref{C45_314} left) and C5 (figure \ref{C45_314} right). As expected, gravitational scattering becomes weaker for C5, but it is still strongly dominating QCD. The cross section for black hole formation has decreased much more than the one for gravitational scattering. 
\par 
Finally also the number of dimensions matters. The model predicts that the more extra dimensions there are the bigger the effects are. Figure \ref{C36_314} gives the angular distributions in the mass bin 3 TeV $< M_{jj} $ TeV for 6 (parameter set C3, plot at the left) and 4 (parameter set C6, plot at the right) extra dimensions. The difference between 6 and 4 extra dimensions is very small. 
\par
Note that when gravitational scattering is the most important contribution, deviations from the QCD cross section are still visible at large $\chi$ values, which is not the case for black holes. But because at large $\chi$ values gravitational scattering is mainly a $t$-channel process, the shape of the graviational scattering curve is close to the QCD one, and the difference between QCD and gravitational scattering disappears in the normalized distributions. 
\par
Deviations from QCD are most visible at low $\chi$ values, and a summary of the above observations is done in table \ref{tabel3}; here we give for a selection of parameter settings and mass bins, the integrated cross sections between $\chi=1$ and $\chi=50$ for QCD ($\sigma_{\mathrm{QCD}}$), gravitational scattering ($\sigma_{\mathrm{GS}}$) and black holes ($\sigma_{\mathrm{BH}}$) separately. 
\begin{table*}
\begin{center}
\caption{Different parameter sets and the relevance of GS and BH in several mass bins.}
\footnotesize
\begin{tabular}{c c c c c c c c c c}
\hline  \hline
 Name & $M_{\mathrm{eff}}$ (TeV)	& n  & $M_s/M_{P}$ & $M_{P}$  &  $M_s$  & Mass bin  & $\sigma_{\mathrm{QCD}}$ & $\sigma_{\mathrm{GS}}$ (nb) & $\sigma_{\mathrm{BH}}$  \\
     &                          &    &            &    (TeV)  &  (TeV)  &  (TeV)   &   (nb)                 & (nb)                      & (nb)           \\
\hline
C1 & 1.0 & 6 & 1.0 & 0.282 & 0.282 &$[1,2]$ & 7.23  & 8.23$\:10^{-3}$ & 113  \\
C2 & 1.0 & 6 & 2.0 & 0.564 & 1.128 &$[1,2]$ & 7.23  & 4.74$\:10^{-2}$ &  20.5\\
C3 & 1.0 & 6 & 4.0 & 1.128 & 4.513 &$[1,2]$ & 7.23  & 7.01$\:10^{-1}$  & 1.01$\:10^{-1}$ \\
C3 & 1.0 & 6 & 4.0 & 1.128 & 4.513 &$[3,14]$ & 2.30$\:10^{-2}$  &8.80$\:10^{-1}$ & 2.33$\:10^{-1}$ \\
C4 & 0.5 & 6 & 8.0 & 1.128 & 9.027 &$[0.5,1]$ & 47.0 & 3.44 & 2.78$\:10^{-3}$ \\
C4 & 0.5 & 6 & 8.0 & 1.128 & 9.027 &$[1,2]$ & 7.23 & 37.37   & 0.10 \\
C4 & 0.5 & 6 & 8.0 & 1.128 & 9.027 &$[2,3]$ & 6.96$\:10^{-2}$ & 2.98  & 1.22$\:10^{-1}$ \\
C4 & 0.5 & 6 & 8.0 & 1.128 & 9.027 &$[3,14]$ & 2.30$\:10^{-2}$ & 1.89  & 2.33$\:10^{-1}$ \\
C5 & 1.0 & 6 & 8.0 & 2.257 & 18.05 &$[3,14]$ & 2.30$\:10^{-2}$ &5.43$\:10^{-1}$ & 4.15$\:10^{-3}$ \\
C6 & 1.0 & 4 & 4.0 & 1.263 & 5.053 &$[3,14]$ & 2.30$\:10^{-2}$ & 8.15$\:10^{-1}$ & 1.45$\:10^{-1}$ \\
 \hline
\end{tabular}
\label{tabel3}
\end{center}
\end{table*}

We will now compare the different new physics scenarios with QCD for a given integrated luminosity, so that we can establish a discovery potential. 
We use the distributions that are normalized to $\chi<50$ and perform a chi-square ($\chi^2$) test between them. 

We have used the following recipe.
For a given integrated luminosity, we consider the normalized distributions $(dN_{\mathrm{QCD}}/d\chi)/N_{\mathrm{QCD}}$ and  $(dN_{\mathrm{total}}/d\chi)/N_{\mathrm{total}}$, with $N_{\mathrm{QCD}}$ and  $N_{\mathrm{total}}$ respectively the number of QCD and total (= QCD + GS + BH) events.
We perform a chi-square ($\chi^2$) test between these distributions to test the null hypothesis that $(dN_{\mathrm{total}}/d\chi)/N_{\mathrm{total}}$ follows the QCD distribution.
We use both a statistical and systematic uncertainty for the calculation of $\chi^2$:
\begin{equation}
\label{chi-square}
\chi^2 = \sum_{\mathrm{all}\:\mathrm{bins}\:i}[\frac{ (\frac{N_{\mathrm{QCD},i}}{N_{\mathrm{QCD}}}-\frac{N_{\mathrm{total},i}}{N_{\mathrm{total}}})^2}{s²_{\mathrm{stat},i}+s²_{\mathrm{sys},i}} ],
\end{equation}
where $N_{\mathrm{QCD},i}$ and $N_{\mathrm{total},i}$ are the number of QCD and total events respectively in bin $i$. The statistical error $s_{\mathrm{stat},i}$ is for each bin taken as $\sqrt{N_{\mathrm{QCD},i}}$. Based on the fact that the theoretical uncertainty does not exceed 20$\%$ (see figure \ref{comErrors} right), and that experimental uncertainties reported by the Tevatron experiments are less than 11$\%$ \cite{PhysRevLett.77.5336,PhysRevLett.80.666}, the systematic error $s_{\mathrm{sys},i}$ is taken to be 25$\%$ over the whole $\chi$ range.
Using $\chi^2$ (equation (\ref{chi-square})) and the number of degrees of freedom, the probability $P$ of having  $dN_{\mathrm{total}}/d\chi/N_{\mathrm{total}}$ given the null hyptothesis is true, can be calculated. The null hypothesis of identity is rejected for $P<0.1$.
\begin{figure*}[ht!]
\begin{center}
\includegraphics[angle=-90,width=0.7\textwidth]{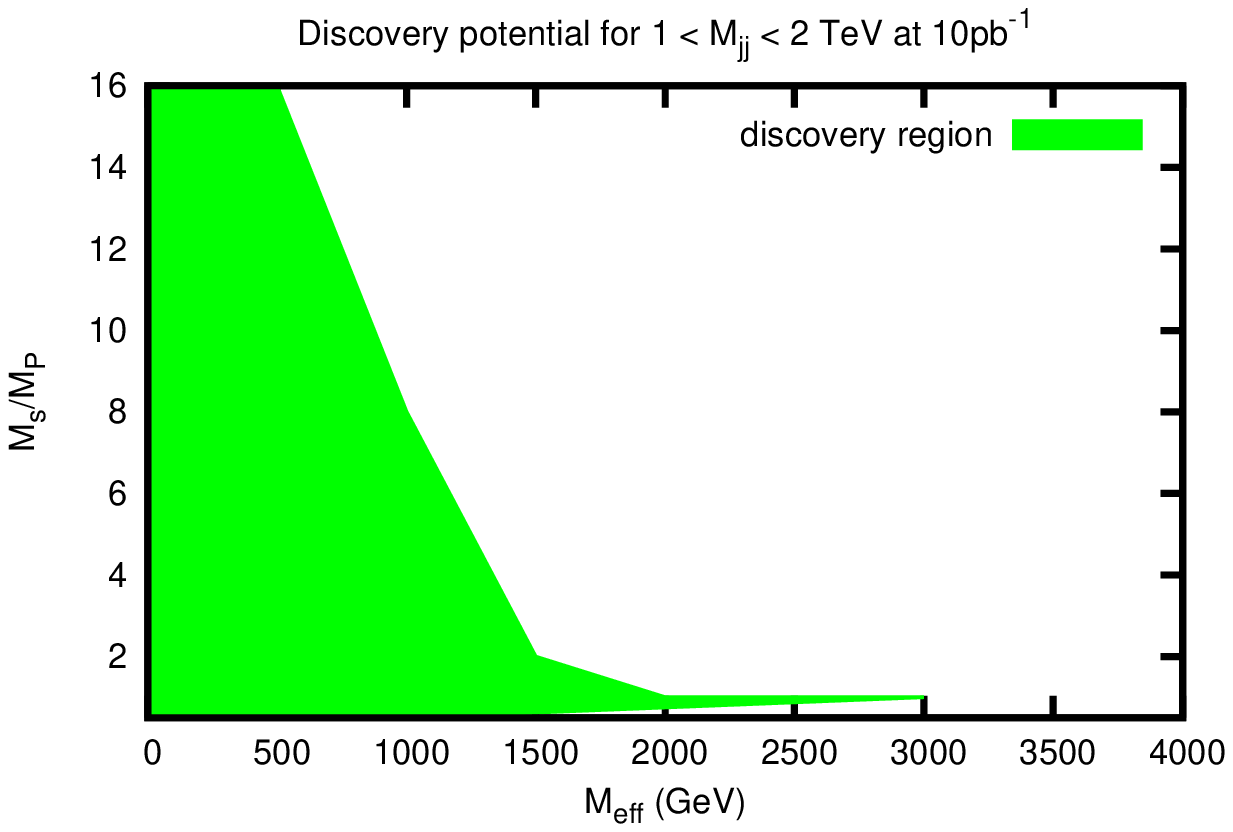}
\caption{Discovery potential for the mass bin $1 < M_{jj} < 2$ TeV at $10pb^{-1}$, assuming a 20$\%$ systematic uncertainty}
\label{1Mjj2_n6_gravity}
\end{center}
\end{figure*}

Let us focus on the mass bin  $1 < M_{jj} < 2$ TeV and consider only those worlds with n=6 extra dimensions and work with an integrated luminosity of 10 pb$^{-1}$. We have used the recipe mentioned above to test several physics scenarios and the green coloured area in figure \ref{1Mjj2_n6_gravity} shows for which model parameters ---$M_{\mathrm{eff}}$ and $M_s/M_{P}$--- the null hypothesis is rejected. This region is from now on called the region of discovery.

As can be seen from the figure, large values of $M_{\mathrm{eff}}$ ($M_{\mathrm{eff}} >$  1.5 TeV) and small values of $M_s/M_{P}$ ($M_s/M_{P} <$ 1) fall outside the region of discovery. The reason is the absence of black hole creation because in that region the lower limit on the black hole mass is drastically increasing with decreasing $M_s/M_{P}$. See equation (\ref{BH_mass1}) and the discussion underneath.
\par
\begin{figure*}
\begin{center}
\includegraphics[angle=-90,width=0.48\textwidth]{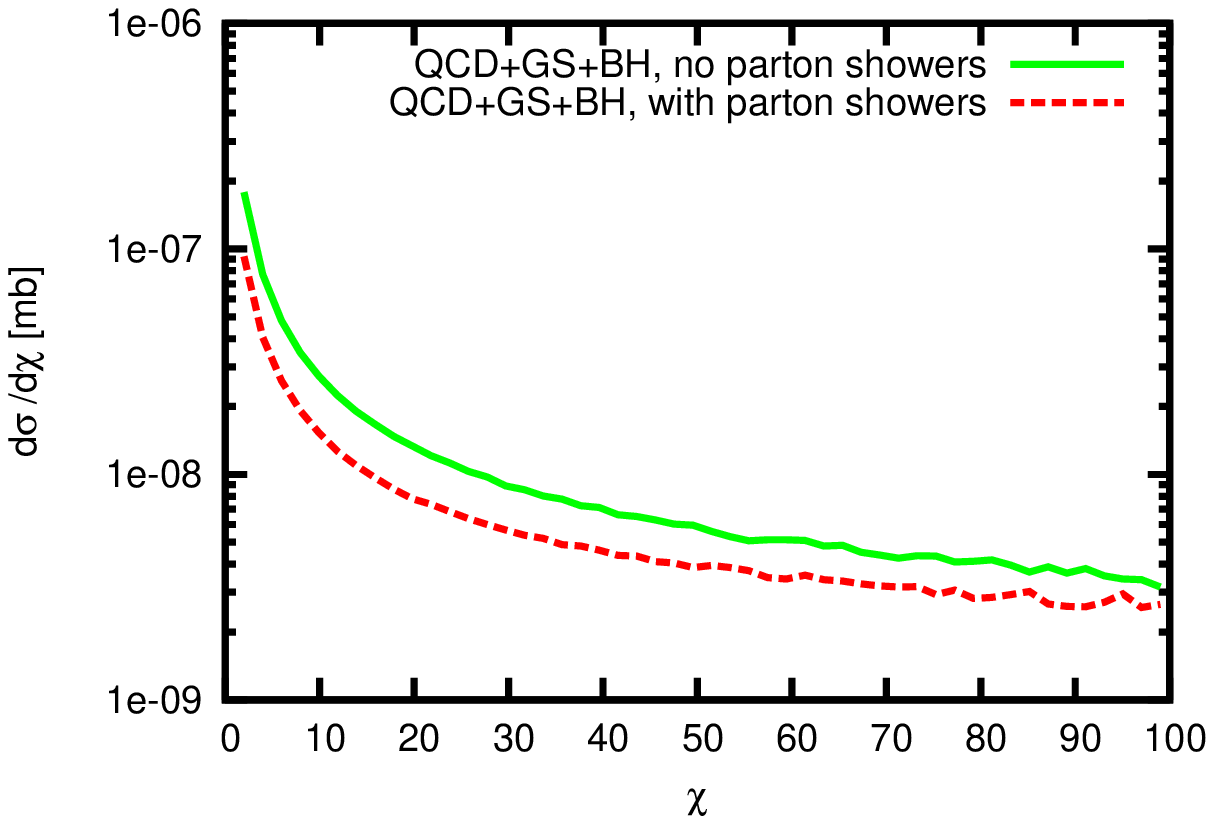}
\hfill
\includegraphics[angle=-90,width=0.48\textwidth]{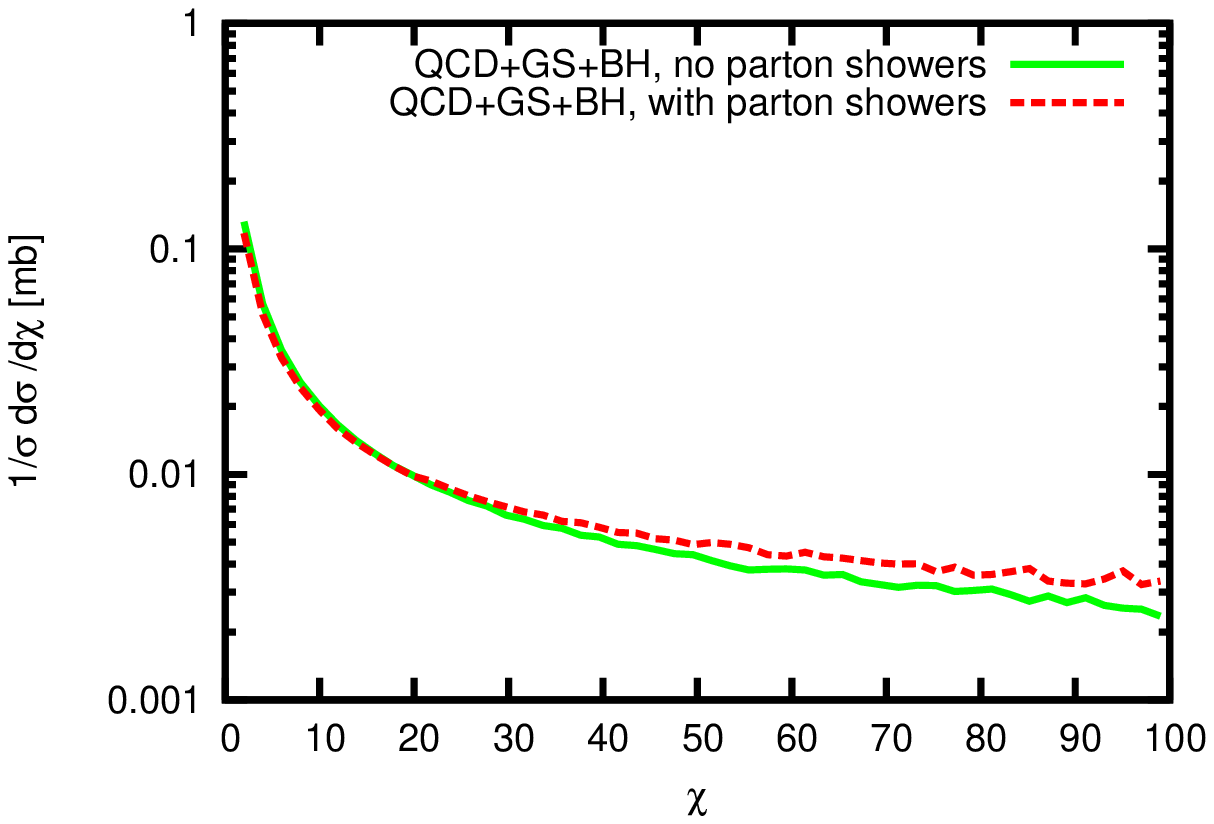}
\caption{Angular distributions in the mass bin 3 TeV $< M_{jj}$ for C3, with and without parton showers. Left: cross section in mb, right: cross section normalized to unit area $1< \chi < 100$.}
\label{showers}
\end{center}
\end{figure*}

Above calculations have been done without parton showering, multiple interactions or hadronization. In figure \ref{showers} we compare the angular distributions with and without parton showers in the mass 3 TeV $< M_{jj}$ bin for C3. The effect of parton showers is most visible at low $\chi$ values.

\section{Conclusions}
\label{conclu}
We have discussed dijet angular distributions at $\sqrt{s}=14$ TeV. First we have performed a QCD study, then we have shown the distributions in a scenario with gravitational scattering and black hole formation in large extra dimensions.
\par
We have used two different programs, JETRAD and NLOJET++, for the calculation of QCD up to next-to-leading order, and found that both programs generate similar results. The angular distributions at NLO are flatter than the Born calculations, especially at large values of $\chi$  ($\chi>100$), which is mainly caused by the fact that the running of $\alpha_s$ has less effect on NLO than on LO calculations.  
Different jet algorithms tend to keep the shape of the distributions unchanged, but give a different normalization.

We have investigated the systematic uncertainties coming from the choice of renormalization  ($\mu_R$)  and factorization  ($\mu_F$) scale and parton distribution function (PDF), and found that a change in $\mu_F$ and PDF mainly influences the normalization. On the other hand, a change in $\mu_R$ has an impact on both the normalization and the shape of the distributions; the distributions have a similar behavior at low  $\chi$ but tend to spread out as $\chi$ increases. 
\par
The effects on dijet angular distributions from gravitational scattering and black hole production, have been studied in the ADD scenario, with the extra requirement that the membrane on which the standard model fields are allowed to propagate, has a finite but small width. The model parameters are the fundamental Planck scale, the width of the membrane and the number of extra dimensions, and it has turned out that the phenomenology is very much dependent on the fundamental Planck scale and the width of the membrane. For a fundamental Planck scale of around 1 TeV and for a wide range of parameter settings, quantum gravity effects have been observed in mass bins above 1 TeV.  For small widths of the membrane, gravitational scattering is the most important process, while black hole formation dominates for wider membranes. In both cases, the effects mainly show up at small values of  $\chi$. The same conclusions can be made for the normalized distributions. Using the shape of the distributions, rather than the absolute normalization, for $\chi<50$, we have determined the region of parameter space that could be discovered with 10 pb$^{-1}$ integrated luminosity and a 25$\%$ systematic uncertainty in the mass bin $1 < M_{jj} < 2$ TeV.
\par 
In conclusion, uncertainties from QCD that cannot be reduced by normalizing the distributions mainly show up at large values of $\chi$ ($\chi>100$), while effects from quantum gravity are mostly present at small values of $\chi$. Depending on the region of $\chi$ under study, dijet angular distributions can therefore either be used as a probe for new physics or as a test of QCD.

\begin{acknowledgement}
The authors wish to thank Torbj\"orn Sj\"ostrand and Leif L\"onnblad for the many useful discussions.
\end{acknowledgement}


\begin{thebibliography}{unsrt}

\bibitem{UA1}
UA1 Collaboration: G.~Arnison  {\it et al.}, Phys. Lett., {\bf B136} (1984) 294;
G.~Arnison  {\it et al.}, Phys. Lett., {\bf B158} (1985) 494;  
G.~Arnison  {\it et al.}, Phys. Lett., {\bf B177} (1986) 244.

\bibitem{UA2}
UA2 Collaboration: P.~Bagnaia {\it et al.}, Phys. Lett., {\bf B144} (1984) 283.

\bibitem{PhysRevLett.77.5336}
CDF Collaboration: F.~Abe {\it et al.}, Phys. Rev. Lett., {\bf 62} (1989) 3020; 
Phys. Rev. Lett., {\bf 69} (1992) 2896; 
Phys. Rev. Lett., {\bf 77} (1996) 5336;
CDF/ANAL/JET/PUB/9609 (2008).

\bibitem{PhysRevLett.80.666}
D0 Collaboration: B.~Abbott {\it et al.}, Phys. Rev. Lett., {\bf 80} (1998) 666--671; 
V.M.~Abazov  {\it et al.}, Phys. Rev. Lett., {\bf 103} (2009) 191803.

\bibitem{atlas}
ATLAS Collaboration: \textit{ATLAS detector and physics performance} (CERN, Geneva 1999). 

\bibitem{giele-1993-403}
W.~T. Giele, E.~W. Glover and D.~A. Kosower, Nucl. Phys., {\bf B403} (1993) 633.

\bibitem{Nagy:2003tz}
Z.~Nagy, Phys. Rev., {\bf D68} (2003) 094002.

\bibitem{catani-1998-510}
S.~Catani and M.~H. Seymour, Nucl. Phys., {\bf B485} (1997) 291--419. 

\bibitem{Catani1993187}
S.~Catani, Y.~L. Dokshitzer, M.~H. Seymour and B.~R. Webber, Nucl. Phys., {\bf B406} (1993) 187--224.

\bibitem{PhysRevD.48.3160}
S.~D. Ellis and D.~E. Soper, Phys. Rev., {\bf D7} (1993) 3160--3166.

\bibitem{siscone}
G.~P. Salam and G.~Soyez, JHEP, {\bf 0705} (2007) 086.

\bibitem{UA1-Cone}
UA1 Collaboration: G.~Arnison {\it et al.}, Phys. Lett. {\bf B 132}, (1983) 214.

\bibitem{RunII}
G.~C. Blazey {\it et al.}, arXiv:hep-ex/0005012 (2000).


\bibitem{pumplin-2002-0207}
J.~Pumplin, D.~R. Stump, J. Huston, H.~L. Lai, P. Nadolsky and W.~K. Tung, JHEP, {\bf 0207} (2002) 012.

\bibitem{nadolsky-2008}
P.~M. Nadolsky, H.-L. Lai, Q.-H. Cao, J.~Huston, J.~Pumplin, D.~Stump, W.-K. Tung and C.-P. Yuan, Phys. Rev., {\bf D78} (2008) 013004.


\bibitem{martin-2009}
A.~D. Martin, W.~J. Stirling, R.~S. Thorne and G.~Watt, arXiv:0901.0002 [hep-ph] (2009).

\bibitem{LHCPrimer}
J.~M. Campbell, J.~W. Huston and W.~J. Stirling, Rep. Prog. Phys., {\bf 70} (2007) 89.  


\bibitem{arkanihamed-1998-429}
N.~Arkani-Hamed, S.~Dimopoulos and G.~Dvali, Phys. Lett., {\bf B429} (1998) 263.

\bibitem{arkanihamed-1999-59}
N.~Arkani-Hamed, S.~Dimopoulos and G.~Dvali, Phys. Rev., {\bf D59} (1999) 086004.

\bibitem{kapner-2007-98}
D.~J. Kapner  {\it et al.}, Phys. Rev. Lett., {\bf 98} (2007) 021101. 

\bibitem{lonnblad-2006-0610}
L.~L\"onnblad and M.~Sj\"odahl, JHEP, {\bf 0610} (2006) 088. 

\bibitem{han-1999-59}
T.~Han, J.~D. Lykken and R.-J. Zhang, Phys. Rev., {\bf D59} (1999) 105006.

\bibitem{giudice-1999-544}
G.~F. Giudice, R.~Rattazzi, and J.~D. Wells, Nucl. Phys., {\bf B544} (1999) 3--88.

\bibitem{sjodahl-2006}
M.~Sj\"odahl and G.~Gustafson, EPJ, {\bf C53} (2008) 109.

\bibitem{Myers1986304}
R.~C. Myers and M.~J. Perry, Annals of Physics, {\bf 172} (1986) 304--347.

\bibitem{1126-6708-2005-09-019}
L.~L\"onnblad, M.~Sj\"odahl, and T.~\AA kesson, JHEP, {\bf 0509} (2005) 019.

\bibitem{pythia}
T.~Sj\"ostrand, S.~Mrenna and P.~Skands, JHEP, {\bf 0605} (2006) 026.

\bibitem{harris-2003-0308}
C.~M. Harris, P.~Richardson and B.~R. Webber, JHEP, {\bf 0308} (2003) 033.


\end{thebibliography}

\end{document}